\documentclass[a4paper,11pt]{article}
\usepackage[utf8]{inputenc}
\usepackage[T1]{fontenc}
\usepackage{amsmath}
\usepackage{amsfonts}
\usepackage{amssymb}
\usepackage[english]{babel}
\usepackage{float}
\usepackage{graphicx}
\usepackage[rmargin=2.5cm,lmargin=2.5cm,tmargin=3cm]{geometry}
\usepackage{lmodern}
\usepackage{url}
\usepackage{hyperref}
\usepackage{lastpage}
\usepackage{lettrine}
\usepackage{color}
\usepackage{tikz}
\usepackage{tikzscale}
\usetikzlibrary{calc, arrows, backgrounds, trees}
\usepackage{pdflscape}
\usepackage[nottoc]{tocbibind}
\usepackage{booktabs}
\usepackage{arydshln}
\usepackage[export]{adjustbox}
\usepackage{float}
\usepackage{rotating}
\usepackage{nameref}
\usepackage{stmaryrd}
\usepackage{verbatim}
\usepackage{color, colortbl}
\usepackage{makecell}
\usepackage{amsmath}
\usepackage{amsthm}
\usepackage{mathtools}

\usepackage{natbib}
\setcitestyle{aysep={}} 
\usepackage{url}

\newtheorem{theorem}{Theorem}
\newtheorem{lemma}{Lemma}
\newtheorem{proposition}{Proposition}

\graphicspath{{Figures/}}       			% Specifies subfolder for figures.

\usepackage[linesnumbered,ruled,vlined]{algorithm2e}
\SetKwInput{KwInput}{Input}                % Set the Input
\SetKwInput{KwOutput}{Output}              % set the Output

\usepackage{comment}
\usepackage{caption}
\usepackage{subcaption}
\usepackage{pdflscape}
\usepackage{pdfpages}
\newcommand{\rev}[1]{{#1}}
\usepackage{setspace}
\allowdisplaybreaks

\usepackage{authblk}
\newcommand{\keywords}[1]{{\small{\noindent \bfseries Keywords: }#1}} 

\title{The Storage Location Assignment and Picker Routing Problem: A Generic Branch-Cut-and-Price Algorithm}
\date{}
\author[1,2]{Thibault Prunet}
\author[1]{Nabil Absi}
\author[3]{Diego Cattaruzza}
\affil[1]{\small Mines Saint-Etienne, Univ. Clermont Auvergne, INP Clermont Auvergne, CNRS UMR 6158 LIMOS, F-13120 Gardanne, France, {absi@emse.fr}}
\affil[2]{\small CERMICS, Ecole des Ponts, Marne la Vallée, France, {thibault.prunet@enpc.fr}}
\affil[3]{\small CRIStAL Centre de Recherche en Informatique Signal et Automatique de Lille, University Lille, CNRS, Centrale Lille, Inria, UMR 9189, F-59000 Lille, France, {diego.cattaruzza@centralelille.fr}}

\begin{document}

\maketitle

\begin{abstract}
\noindent
The \textit{Storage Location Assignment Problem} (SLAP) and the \textit{Picker Routing Problem} (PRP) have received significant attention in the literature due to their pivotal role in the performance of the \textit{Order Picking} (OP) activity, the most resource-intensive process of warehousing logistics. The two problems are traditionally considered at different decision-making levels: tactical for the SLAP, and operational for the PRP. However, this paradigm has been challenged by the emergence of modern practices in e-commerce warehouses, where storage decisions are more dynamic and are made at an operational level. This shift makes the operational integration of the SLAP and PRP pertinent to consider. Despite its practical significance, the joint optimization of both operations, called the \textit{Storage Location Assignment and Picker Routing Problem} (SLAPRP), has received limited attention. Scholars have investigated several variants of the SLAPRP, proposing both exact and heuristic methods, including different warehouse layouts and routing policies. Nevertheless, the available computational results suggest that each variant requires an ad hoc formulation. Moreover, achieving a complete integration of the two problems, where the routing is solved optimally, remains out of reach for commercial solvers, even on trivial instances. 

In this paper, we propose an exact solution framework that addresses a broad class of variants of the SLAPRP, including all the previously existing ones. This paper proposes a Branch-Cut-and-Price framework based on a novel formulation with an exponential number of variables, which is strengthened with a novel family of non-robust valid inequalities. We have developed an ad-hoc branching scheme to break symmetries and maintain the size of the enumeration tree manageable. Computational experiments show that our framework can effectively solve medium-sized instances of several SLAPRP variants and outperforms the state-of-the-art methods from the literature.

\end{abstract}


%\tableofcontents

\section{Introduction}

% intro on warehousing logistics: key element of a supply chain, order picking, mainly manual (ref in silva). context picker to part warehouse.
In supply chain management, the storage of goods in warehouses acts as a key component of the system performances~\citep{boysen_warehousing_2019}. The order picking activity (i.e., the action of retrieving the products from their storage locations) is largely considered the most resource-intensive activity in warehousing operations. According to~\cite{de_koster_design_2007}, manual picker-to-parts warehouses, in which a human operator walks through the shelves to collect the products, were still largely dominant in 2007 and accounted for more than 80\% of all warehouses in Western Europe. In such warehouses, the order picking process alone accounts for 50-75\% of the total operating cost~\citep{frazelle_world-class_2016}, leading to a prolific research stream on the optimization of these systems \citep{de_koster_design_2007}.

% this part in the justification of more oeprational problem ?

% short intro on SLAP and OPP (or picker routing ?), usually solved separately -> in practice it translates to a limited information in the SLAP.
There are \rev{three} main decision problems related to the operational efficiency of an order picking system: the Storage Location Assignment Problem (SLAP), which aims at determining an efficient assignment of products, called Stock Keeping Units (SKUs) to storage locations, \rev{the Order Batching Problem (OBP), which aims at grouping customer orders into batches retrieved by single routes,} and the Picker Routing Problem (PRP), which consists in finding the most efficient path in the warehouse to pick a given set of products from their locations. \rev{The SLAP and the PRP} are strongly linked, as the PRP determines the picking routes once the product locations are known, and the quality of a SLAP solution is assessed via the computation of the routes followed by the pickers.

\rev{The SLAP, OBP and PRP have been studied extensively in the literature \citep{de_koster_design_2007,gu_research_2007}. Most of the time, these problems are solved independently or sequentially. Recent studies have, however, highlighted the benefits of making assignment, storage, batching and routing decisions in an integrated way in warehousing logistics \citep{van_gils_designing_2018,van_gils_increasing_2018}. The integrated optimization of the OBP and PRP is currently an active research topic \citep{briant_efficient_2020,wahlen_branch-price-and-cut-based_2023}. To deal with the complexity of integrated problems, operations management decisions are generally categorized based on the time horizon they address, namely strategic, tactical, and operational. Hence, the integration of the SLAP and the PRP remains marginal, as the SLAP is usually considered a tactical problem, whereas the PRP is viewed as an operational problem. However, the paradigm shift in e-commerce warehouses increases the demand variability and often requires a high level of responsiveness to ensure customer satisfaction. This translates into shorter lead times for processing and delivering incoming orders \citep{boysen_warehousing_2019,rijal_workforce_2021}. To face these challenges, a common strategy is to divide the storage area into two zones: the reserve area, where products are stacked in high bays, and the smaller forward picking area, where products are stored in easily reachable racks for pickers. Although storage decisions in the reserve area remain tactical, the current trend in e-commerce warehouses is to consider them as operational in the picking area. Indeed, here the inventory quantities are small, and the \rev{picking is organized in waves, with the replenishment of depleted products generally occurring several times per day \citep{de_vries_prioritizing_2014,guo_storage_2021,celik_inventory_2021}.}} %Figure~\ref{fig:waves} illustrates such an organization.} 

\rev{While a complete reoptimization of the storage plan at the operational level may seem unrealistic, the storage-location assignment decisions for replenished items are made several times per day, with complete information on the incoming demand for the upcoming cycles. This allows the explicit consideration of picker routing decisions within the problem. This dynamic storage replenishment is inspired by \rev{real-world applications} of e-commerce warehouses, e.g., in Germany \citep{weidinger_scattered_2018}, Belgium \citep{bahrami_enhancing_2019} and China \citep{guo_storage_2021}. Other applications involving storage decisions at an operational level include the partial reassignment of already stored items, that can happen on a daily basis in modern real-world warehouses \citep{kim_slotting_2012,kofler_re-warehousing_2011,li_dynamic_2016}. }
%When considered at the tactical level, the joint optimization of the two problems may still be a reasonable approximation of the reality, for instance in intermediate warehouses with recurrent orders for parts used in cyclic production for a low-variety make-to-stock supply chain. 

\rev{In this context of the forward picking area, three decisions are taken before each picking cycle: (1) where to store the SKUs with insufficient inventory to meet the incoming demand (storage), (2) how to group customer orders into consolidated batches retrieved in a single route (batching) and (3) how to route the pickers (routing). Consequently, different alternatives may be considered:}
\begin{enumerate}
    \item \rev{Solve the three decision problems sequentially.}
    \item \rev{First, consider storage with no information on the batches. Then, jointly optimize batching and routing with perfect information on the SKU locations.}
    \item \rev{First, consider batching with incomplete information on the SKU locations. Then, once all batches are known, jointly optimize storage and routing with perfect information on the SKUs retrieved by each route.}
    \item \rev{Jointly optimize storage location, batching, and routing.}
\end{enumerate}

\rev{As already mentioned, the first option has been proved to lead to solutions of poor quality and is therefore discarded \citep{van_gils_increasing_2018}. In the second option, no information on the batching decisions is available to the decision-maker when deciding storage assignment, whereas in the third case, batching is determined with partial, but still rather complete information on the storage, as only a portion of the SKUs are repleted in each replenishment cycle. Finally, it is clear that a complete integration of the three problems would lead to the best results. However, it is worth noting that this integrated approach poses significant challenges and is left for future research studies.}

\rev{In this paper, we assume that batching decisions are made beforehand, and we focus on the joint optimization of the storage and picker routing decisions within the forward picking area, where both decisions may be taken with detailed knowledge of the incoming demand. It is important to note that the orders we use as input data do not correspond to individual customer orders, but to consolidated batches of orders, each retrieved by a single route. The resulting problem, coined the \textit{Storage Location Assignment and Picker Routing Problem} (SLAPRP), has been recently introduced in the literature and has been proven NP-hard in the strong sense, even for very basic cases \citep{boysen_deterministic_2013}. The study of the SLAPRP is also relevant in the case of a completely stochastic demand, where it can serve as a subproblem in complex solution methods.}

%\textcolor{red}{From the previous paragraph, the part {\em While a complete integration ... left for future studies} should go after the part {\em Consequently, two options remain ...}. To me we have to motivate the pertinence of our problem by convincing the reader that it does make sense to integrate only SLAP and PRP in this context. As it is written, it seems that the main reason why we only integrate SLAP and PRP without batching is because it is too challenging. It may be the case, but we should not put stuff in this way.}

% organization of the paper
The remainder of this paper is organized as follows. Section~\ref{sec:literature} presents the context, related background, and the contributions of the current work. %, highlighting its main contributions, with a brief presentation of the related literature. 
Section~\ref{sec:description_and_formulation} defines the SLAPRP and introduces a generic compact formulation. An extended formulation is then derived from a Dantzig-Wolfe reformulation. Section~\ref{sec:og_cuts} introduces a novel family of valid inequalities for the SLAPRP. Section~\ref{sec:algorithm} presents the components of the generic Branch-Cut-and-Price algorithm we designed to solve the problem. Section~\ref{sec:pricing} describes the labeling algorithm used to solve the pricing problems. Section~\ref{sec:experiments} illustrates the numerical experiments. Finally, Section~\ref{sec:conclusion} concludes this work.

\section{Literature Review}\label{sec:literature}

In this section, we present the literature on warehousing problems related to our work. For readability and brevity, we first provide a concise introduction to the PRP and the SLAP, then we review the SLAPRP literature. For a detailed literature review on the PRP and the SLAP, interested readers are referred to \cite{de_koster_design_2007,gu_research_2007,van_gils_designing_2018,reyes_storage_2019,boysen_warehousing_2019}. 

\paragraph{Picker Routing Problem.}
The PRP takes as input the warehouse layout and the location of the different SKUs, it aims at optimizing the material handling time to pick all the SKUs from a pick list, often expressed as the traveled distance. The classical PRP is a special case of the Traveling Salesman Problem (TSP). 
%The nodes represent the drop off point and the storage locations with (at least) one item from the pick list, the vertices represent the shortest distance between two locations. 
\rev{Despite being NP-hard in general \citep{prunet2023intractability},} the problem is solvable in polynomial time for the single-block warehouse \citep{ratliff_order-picking_1983} and the two-blocks warehouse \citep{roodbergen_routing_2001}. For the multiple-blocks warehouse, the problem is fixed-parameter tractable \citep{pansart_exact_2018}. However, most practical variants of the problem remain NP-hard and few works addressed it with exact solution methods \citep{schiffer_optimal_2022}. Nevertheless, the recent work of \citep{cambazard_fixed-parameter_2018} on the rectilinear TSP enabled the development of efficient exact algorithms for the PRP, which solve industrial size instances in a reasonable computing time \citep{pansart_exact_2018,schiffer_optimal_2022}.

\begin{figure}[h!]
    \centering
    \begin{subfigure}{.3\textwidth}
        \includegraphics[width=.95\linewidth]{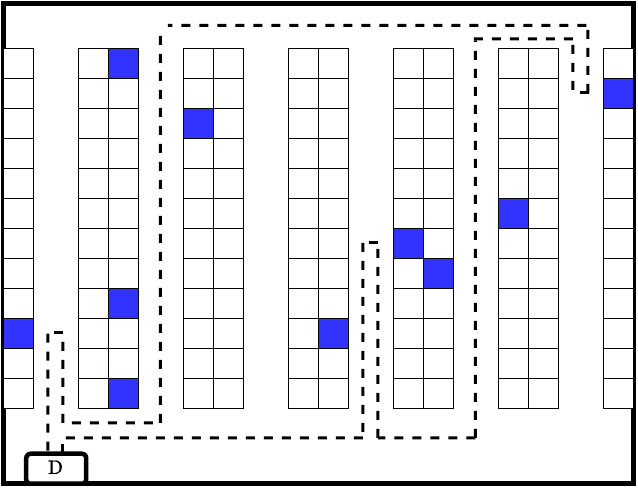}
        \caption{Optimal (58)}
    \end{subfigure}
    \begin{subfigure}{.3\textwidth}
        \includegraphics[width=.95\linewidth]{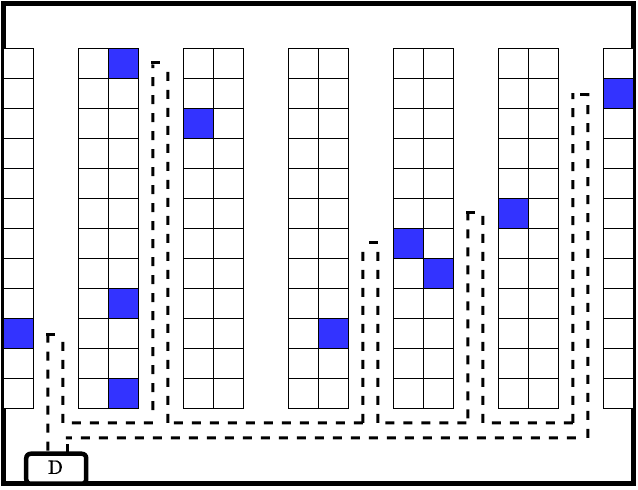}
        \caption{Return (88)}
    \end{subfigure}
    \begin{subfigure}{.3\textwidth}
        \includegraphics[width=.95\linewidth]{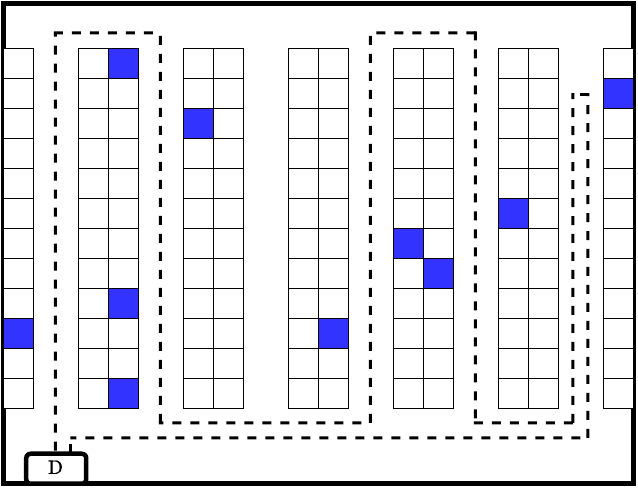}
        \caption{S-shape (84)}
    \end{subfigure}
    \begin{subfigure}{.3\textwidth}
        \includegraphics[width=.95\linewidth]{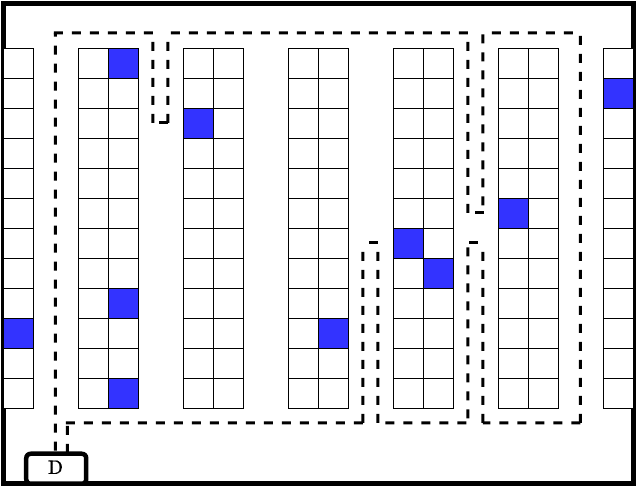}
        \caption{Midpoint (77)}
    \end{subfigure}
    \begin{subfigure}{.3\textwidth}
        \includegraphics[width=.95\linewidth]{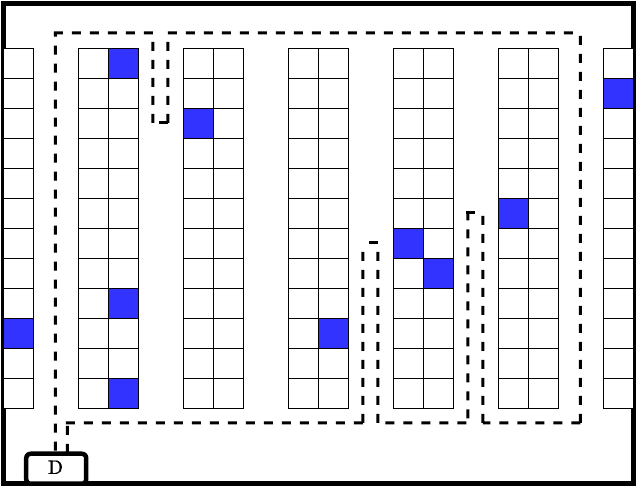}
        \caption{Largest gap (68)}
    \end{subfigure}
    \caption{Routing policies with route distances in brackets}
    \label{fig:routing_policies}
\end{figure}

Many researchers have noted that, despite the availability of competitive PRP algorithms, simple heuristics enjoy much higher popularity \citep{de_koster_design_2007}. The most common of these heuristics, called  \textit{routing policies}, are  presented in Figure~\ref{fig:routing_policies} for a single block warehouse. %These heuristics, called \textit{routing policies}, are usually determined by a set of simple rules. Figure~\ref{fig:routing_policies} presents the most common of these policies for a single block warehouse. 
With the \textit{S-shape policy} the picker traverses completely an aisle if there is an item to pick. A return is allowed from the last picking location, so that the last aisle may not be traversed entirely. In the \textit{return policy}, the back cross aisle is never used: In each aisle, the picker goes up to the farthest picking location and then comes back to the front cross aisle. The \textit{midpoint policy} considers that the first and last aisles containing an SKU to pick are traversed completely. In the other aisles the picker never crosses the middle of the aisle: The picking locations above the midpoint are visited coming from the top cross aisle, while the ones below the midpoint are visited coming from the bottom cross aisle. The \textit{largest gap policy} is similar to the midpoint policy, except that the position of the \lq\lq midpoint" (i.e., the point that is never crossed in an aisle) may not be located in the middle of the aisle, but in the location minimizing the within-aisle distance.

%%%%%%%%%%%%%%%%%%%%%%%%%%%%%%%%%%%%%%%%%%%%%%%%%%%%%%%%%%%%%%%%%
\paragraph{Storage decisions.}

% intro on the slap
The SLAP was introduced by \cite{hausman_optimal_1976} for an automated warehouse and most of its variants are NP-hard \citep{reyes_storage_2019}. %and has been studied extensively by the community since then \citep{gu_research_2007}. In terms of structure, the SLAP belongs to the family of assignment problems, an important class of problem in Operations Research. It is especially related to the Quadratic Assignment Problem (QAP), considered as one of the most challenging combinatorial optimization problem due to its non-linear objective function \citep{loiola_survey_2007}. It is then not surprising that most of the SLAP variants are NP-hard \citep{reyes_storage_2019}. 
Various performance evaluation measures have been used for the SLAP, the most common are the total picking time and traveled distance by the pickers \citep{reyes_storage_2019}. Note that, given a solution of the SLAP (i.e., a storage plan), the modeling and computation of the picking distance largely differ between different works, %, as the issue is not straightforward: the exact walking distance is a complex function that depends on the warehouse layout, the chosen routing policy, the characteristics of the products, and the product/location assignment itself. 
and in most cases, a distance-related indicator is used as the objective function instead of the distance itself.

% classical storage policies
The SLAPs are classically solved with simple heuristics called \textit{storage policies} that assign the incoming items to open storage locations according to a simple rule. The taxonomy introduced by \cite{hausman_optimal_1976}, still widely used today, separates these policies into three classes: i. The \textit{random storage} policy in which an incoming item is allocated to a random open storage location. %This policy is adapted to the case where no information is known on the incoming products. 
%Despite its simplicity of implementation, its major drawback is to not optimize the picking distance. 
ii. The \textit{dedicated storage} policy, where each SKU is assigned to a single storage location in the warehousing area. The products are sorted according to a given criterion, usually the turnover, and assigned to the closest storage locations accordingly. iii. The \textit{class-based storage} policy assigns the products to \lq\lq classes" depending on a frequency-related criterion. Each class is assigned to a zone in the warehouse. Within a zone, the specific location of a given product is randomly determined.

\paragraph{Interactions between the SLAP and the PRP.}

The two problems are closely linked. On the one hand, a SLAP solution must be known \textit{a priori} when solving a PRP instance. On the other hand, a  PRP must be solved to assess \textit{a posteriori} the quality of a SLAP solution. It is therefore not surprising that the two problems have been studied together giving rise to the SLAPRP. %: In light of the popularity of storage and routing policies to tackle independently the SLAP and the PRP, a natural question arises concerning which routing policy works best for a given storage policy, and vice versa. Several studies have tackled this issue by solving the two problems sequentially, under different circumstances, e.g. \citep{petersen_comparison_2004,roodbergen_simultaneous_2015,van_gils_increasing_2018} among others. 

\paragraph{The SLAPRP.}
When the information is known \textit{a priori} at the order level, i.e., not only the demand for each product but the list of products in every order, the SLAP and the PRP can be solved together. The SLAPRP is the integrated variant of these two problems, which aims at finding an assignment plan that minimizes the total picking distance, computed exactly by solving a PRP for each order. The integration of location and routing decisions is not a novel idea \citep{van_gils_increasing_2018,van_gils_designing_2018}, yet few works have studied the SLAPRP. These two problems are indeed classically seen as belonging to different decision levels, the SLAP being tactical and the PRP operational. However, the recent rise of e-commerce warehouses has led to a shift in this paradigm to better adapt to versatile demand with a high level of responsiveness, toward a more dynamic picking process where the storage decisions are taken more frequently, at the operational level. This is especially true for the problem variants that focus on the reorganization and replenishment of the picking area, which take place several times per day in modern warehouses \citep{guo_storage_2021,weidinger_scattered_2018,bahrami_enhancing_2019}. Recent survey papers on integrated planning problems in warehousing \citep{van_gils_designing_2018}, and e-commerce warehouses \citep{boysen_warehousing_2019}, highlight the need to integrate decision problems to improve the efficiency of order picking.

The first example of a SLAPRP encountered in the literature is the work of \cite{mantel_order_2007} that studies an automated warehouse with a vertical lift system. The authors propose a MIP formulation of the studied problem with the S-shape routing policy. Their problem is a very specific instance of the SLAPRP: With the S-shape routing, each aisle is traversed completely and the model only needs to decide in which aisle a product is located, leading to a much simpler problem. \cite{boysen_deterministic_2013} study the SLAPRP with the return routing policy on a simple layout with only a single aisle. They prove that the problem is NP-hard in the strong sense, even in simple cases, and provide a dynamic programming algorithm to solve it. \citep{silva_integrating_2020} is the first work, to our best knowledge, that studies a fully integrated SLAPRP where the PRP is solved to optimality, instead of using a routing policy. They propose a cubic formulation for the SLAPRP with the optimal routing, as well as linear, quadratic, or cubic formulations for other widespread routing policies (return, S-shape, midpoint, and largest gap). They linearize their formulations for the computational experiments with commercial solvers. The results show that the SLAPRP with optimal routing is significantly more challenging than its variants with the heuristic routing policies. Only trivial instances are solved to optimality. They also propose a Variable Neighborhood Search heuristic to tackle larger instances. Finally, \cite{guo_storage_2021} study a variant of the SLAPRP focusing on the replenishment of the picking area\rev{, based on an industrial case of a Chinese third-party logistic company.} In their problem, a portion of the products are already in place in the picking area, and only a set of new items needs to be assigned to the open storage locations. %Their problem focuses on a specific warehouse layout, provided by the studied case company. 
They propose a MIP formulation with the return routing policy, and a dynamic programming-based heuristic to solve the problem.

\paragraph{Paper contributions.}
In the light of this literature review, it appears that the study of the SLAPRP answers to modern industrial practices, yet the related literature is scarce, with most existing studies focusing on specific application cases. Only one work \citep{silva_integrating_2020} studies the problem with optimal routing instead of heuristic policies. In this paper, we aim at designing an exact solution method for a large class of variants of the SLAPRP, namely, the fully integrated problem, its variants with the classical routing policies (return, S-shape, midpoint, and largest gap), and the replenishment variants of these problems. The main contributions of this work are the following:
\begin{enumerate}
    \item We introduce two novel formulations of the SLAPRP: a compact linear formulation for optimal routing and a general extended formulation based on a Dantzig-Wolfe reformulation. This extended formulation is independent of all operational considerations.
    \item We introduce a new family of valid inequality for the extended formulation. Since these inequalities are non-robust, particular attention is given to the management of the associated cutting planes.
    \item We propose a generic Branch-Cut-and-Price algorithm that can solve a large class of SLAPRP variants, including various warehouse layouts and picker routing policies, as well as the partial replenishment of the picking area. The only adaptations from one variant to another are i. the underlying graph, and ii. the resource extension function in the pricing problems.
    \item We present an extensive set of computational experiments that shows the competitiveness of the method in various configurations. 
\end{enumerate}
\section{Problem description and mathematical formulations}\label{sec:description_and_formulation}

In this section, we formally introduce the SLAPRP and propose two generic formulations of the problem, one compact and another one extended. We use the term \textit{generic} to indicate that we do not assume any particular layout on the storage area, nor any particular routing strategy (note that the illustrative example in Figure~\ref{fig:graph_representation:1} refers to a single block rectangular warehouse). Table~\ref{tab:notations} summarizes the notation used throughout this paper.

\subsection{Problem description \&  compact formulations}\label{sec:problem_description}

\begin{figure}[h]
    \centering
    \begin{minipage}[b]{0.45\textwidth}
        \includegraphics[width=\textwidth]{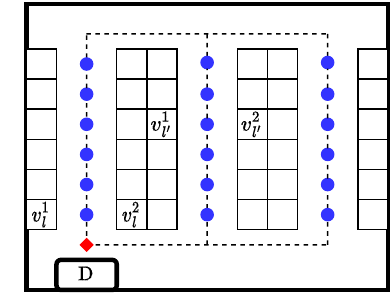}
        \caption{Warehouse layout with storage locations (blue) and drop-off point (red).}
        \label{fig:graph_representation:1}
    \end{minipage}
    \hfill
    \begin{minipage}[b]{0.45\textwidth}
        \includegraphics[width=.9\textwidth]{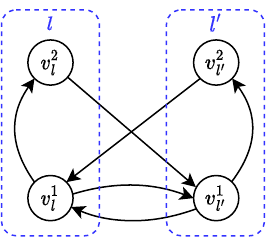}
        \caption{Induced subgraph of $\mathcal{G}$ from nodes $v_l^1$, $v_l^2$, $v_{l'}^1$ and $v_{l'}^2$.}
        \label{fig:graph_representation:2}
    \end{minipage}    
\end{figure}

% explanation of warehouse layout and modeling of it with aggregated locations
% graph representation of the layout -> storage location = aggregation of storage spaces, to reduce the size of our model, and reduce symmetries. The PRP is defined on this graph
% this graph is NOT a Steiner graph
\paragraph{Storage nodes and locations.}
%Let us introduce some concepts to model the layout of the storage area. 
We assume that all picker routes start and end at the same location $v_0$, called the \textit{drop-off point}, and that a picker can indifferently pick from the right and left shelves when crossing an aisle. We call \textit{storage node} an elementary storage space that can hold at most one SKU. We note $\mathcal{V}$ the set of storage nodes, and $\mathcal{V}^0 = \mathcal{V} \cup \{v_0\}$. We call \textit{storage location} a set of storage nodes that are equivalent in terms of distance from the other storage nodes. For example, in Figure~\ref{fig:graph_representation:1}, the storage nodes $v_l^1$ and $v_l^2$ are part of the same storage location $l$ (represented by the blue dot in the middle). We note $\mathcal{L}$ the set of storage locations, $\mathcal{V}(l)$ the set of storage nodes that are part of location $l \in \mathcal{L}$, and $K_l = |\,\mathcal{V}(l)\,|$ its \textit{capacity}. Note that $\mathcal{L}$ forms a partition of $\mathcal{V}$, so we can index the set as $\mathcal{V}=\bigcup_{l \in \mathcal{L}}\{v_l^1, v_l^2,\dots, v_l^{K_l}\}$. For simplicity we can use {\em node} (resp. {\em location}) instead of {\em storage node} (resp. {\em storage location}).

\paragraph{Graph representation.}
The layout of the warehouse can be modeled with a directed graph $\mathcal{G} = (\mathcal{V}^0, \mathcal{E})$ where $\mathcal{E} = \{(v_0, v^1_l): l \in \mathcal{L}\} \cup \{(v^i_l, v_0): l \in \mathcal{L}, i = 1,\dots, K_l\} \cup\{(v^i_l, v^1_{l'}): l, l'\in\mathcal L, i = 1, \dots, K_l\} \cup \{(v^i_l, v^{i+1}_l): l\in\mathcal{L}, i=1,\dots, K_l - 1\}$. The graph is thus not complete: for a location $l \in \mathcal{L}$, the node $v_l^1$ is reachable from all the other nodes in the graph, but for $2 \leq i \leq K_l$ the node $v_l^i$ is only reachable from node $v_l^{i-1}$. In other words, the node $v_l^i$ represents the $i^{th}$ visit of location $l$, which is only reachable after $(i-1)$ previous visits of $l$. Figure~\ref{fig:graph_representation:2} provides a minimal example with two locations and four nodes. The drop-off point $v_0$ is reachable from all the other nodes. Note that the arcs in $\mathcal{E}$ are not associated with a weight in the general case, since the distance between two nodes depends on the routing policy. Actually, for some policies, the distance between two nodes is not constant as it depends on the nodes previously visited by a path. We assume that the distances satisfy the triangle inequality.
\paragraph{Problem description.}
The SLAPRP can be formally described as follows. A graph warehouse layout is given, represented by a graph $\mathcal{G}$ as previously introduced. The aim is to assign a set of SKUs $\mathcal{S}$ to the storage nodes $\mathcal{V}$. Without loss of generality, we assume that $|\mathcal{S}| \leq |\mathcal{V}|$. A set of orders $\mathcal{O}$ needs to be collected by the pickers. An order $o \in \mathcal{O}$ contains a subset $\mathcal{S}(o)$ of SKUs to be picked, and corresponds to one picking route. A route is defined as a tour on graph $\mathcal{G}$ starting and ending at the drop-off point. A route is valid if and only if it visits the storage nodes assigned to its set of SKUs $\mathcal{S}(o)$. Each route is associated with a cost that represents the walking distance covered by a picker and depends on the routing policy. A set $\mathcal{F} \subset \mathcal{S} \times \mathcal{L}$ of \textit{fixed assignments} represents the locations that are already filled. For a pair $(s,l) \in \mathcal{F}$, a SLAPRP solution may only be valid if SKU $s$ is assigned to location $l$.

\paragraph{Compact formulation.}
We introduce two sets of binary decision variables: $\xi_{ls}$ equals 1 if SKU $s \in \mathcal{S}$ is stored into location $l \in \mathcal{L}$, 0 otherwise, and $x_{ij}^o$ equals 1 if the route picking order $o\in\mathcal O$ uses arc $(v_i,v_j)\in\mathcal{E}$, 0 otherwise. We also introduce non-negative continuous variables $u_i^o$ that indicate the position of location $i\in\mathcal{V}$ in the tour picking order $o\in\mathcal{O}$.
Finally, the picking distance of a given route $\underline{x^o} = (x^o_{ij})_{(i,j) \in \mathcal{E}}$ is denoted $f(\underline{x^o})$. Note that we do not make any assumption on the function $f$, which depends on the considered variant of SLAPRP. With these definitions, we propose the following compact formulation of the SLAPRP (denoted $C$).
\begin{align}
    &\textbf{(C)} &\min\, & \sum_{o \in \mathcal{O}} f(\mathbf{\underline{x^o}}) & & \label{eq:c1:obj}\\
    & &    s.t.\, & \sum_{s \in \mathcal{S}} \xi_{ls} \leq K_l   & \forall l \in \mathcal{L} & \label{eq:c1:01} \\
    & &    & \sum_{l \in \mathcal{L}} \xi_{ls} = 1  & \forall s \in \mathcal{S} & \label{eq:c1:02}\\
    & &    & \xi_{ls} = 1 & \forall (s,l) \in \mathcal{F} & \label{eq:c1:022}\\
    & &    & \sum_{v_i \in \mathcal{V}} x_{i0}^o = \sum_{v_j \in \mathcal{V}} x_{0j}^o = 1  & \forall o \in \mathcal{O} & \label{eq:c1:03}\\
    & &    & \sum_{v_j \in \mathcal{V}^0} x_{ij}^o = \sum_{v_j \in \mathcal{V}^0} x_{ji}^o & \forall v_i \in \mathcal{V}, o \in \mathcal{O} & \label{eq:c1:04} \\
    & &    & u^o_j \geq u^o_i + 1 - |\mathcal{S}(o)|\, (1-x_{ij}^o) & \forall o \in \mathcal{O}, v_i,v_j \in \mathcal{V}& \label{eq:c1:05}\\
    & &    & \sum_{v_i \in \mathcal{V}^0} \sum_{v_j \in \mathcal{V}} x_{ij}^o = |\mathcal{S}(o)| & \forall o \in \mathcal{O} & \label{eq:c1:06}\\
    & &    & \sum_{v_i \in \mathcal{V}^0} \sum_{v_j \in \mathcal{V}(l)} x_{ij}^o \geq \sum_{s \in \mathcal{S}(o)} \xi_{ls} & \forall l \in \mathcal{L}, o \in \mathcal{O} & \label{eq:c1:07} \\
    %& &    & \rev{0 \leq f^{os}_{ij} \leq x^o_{ij}} & \rev{\forall o \in \mathcal{O}, s \in \mathcal{S}(o), i,j \in \mathcal{V}^0} & \label{eq:c1:10bis} \\
    & &    & x_{ij}^o \in \{0,1\} & \forall o \in \mathcal{O}, l,h \in \mathcal{V}^0 & \label{eq:c1:08}\\
    & &    & \xi_{ls} \in \{0,1\} & \forall l \in \mathcal{L}, s \in \mathcal{S}&\label{eq:c1:09}\\
    & &    & 0 \leq u^o_i \leq |\mathcal{S}(o)| - 1 & \forall o \in \mathcal{O}, v_i \in \mathcal{V} & \label{eq:c1:10}
\end{align}

The objective function~\eqref{eq:c1:obj} minimizes the total distance walked to collect all orders. Constraints~\eqref{eq:c1:01} and~\eqref{eq:c1:02} are assignment constraints that ensure that each SKU is assigned to a single storage location, and that the number of SKUs stored in one location does not exceed the capacity. Constraints~\eqref{eq:c1:022} enforce the fixed assignments. Constraints~\eqref{eq:c1:03}, \eqref{eq:c1:04} and~\eqref{eq:c1:05} are tour constraints. They ensure that, for each order, the corresponding route forms a valid tour. Constraints~\eqref{eq:c1:03} impose that a route is formed to pick each order, constraints~\eqref{eq:c1:04} enforce the flow conservation at each location, and constraints~\eqref{eq:c1:05} deal with subtour elimination, based on the Miller-Tucker-Zemlin (MTZ) formulation. Constraints~\eqref{eq:c1:06} ensure that the route collecting an order visits exactly as many locations as SKUs to retrieve. These constraints are redundant in this compact formulation, but they prove to be useful once convexified. Constraints~\eqref{eq:c1:08} are linking constraints and ensure that the assignment of SKUs and the chosen routes are consistent. More precisely, for each order $o \in \mathcal{O}$, the route picking $o$ must stop at location $l \in \mathcal{L}$ (left-hand side) if one of the SKUs of the order is assigned to this location. If several SKUs of $o$ are stored in $l$, the route must stop multiple times. Constraints~\eqref{eq:c1:08}-\eqref{eq:c1:10} define the domains of the variables. Note that the main decision variables $\xi$ do not appear in the objective function, supporting the discussion of Section~\ref{sec:literature} on the link between the SLAP and the PRP. 
\rev{\paragraph{Note on the subtour elimination constraints.} The MTZ constraints used in formulation (C) are known to provide formulations with a poor linear relaxation. We present in Appendix~\ref{appendix:compact_mcf} a lifting of (C) that relies on the multi-commodity flow (MCF) formulation for subtour elimination. Although this strengthens the compact formulation, the subtour constraints are expressed using the assignment variables $\xi$, which would be cumbersome to decompose the problem. The numerical experiments presented in Section~\ref{sec:experiments} use both compact formulations.}

%%%%%%%%%%%%%%%%%%%%%%%%%%%%%%%%%%%%%%%%%%
% Notations %
%%%%%%%%%%%%%%%%%%%%%%%%%%%%%%%%%%%%%%%%%

\begin{table}[htp]
\centering
\caption{Mathematical notations}
\label{tab:notations}
\begin{tabular}{ll}
\hline
\multicolumn{2}{l}{Sets}       \\ \hline
$\mathcal{V}$ & Set of elementary storage nodes       \\ 
$\mathcal{V}^0$ & Set of nodes with the drop off point \\
$\mathcal{E}$ & Set of arcs of graph $\mathcal{G}$ \\
$\mathcal{L}$ & Set of storage locations \\
$\mathcal{V}(l)$ & Set of nodes associated with location $l \in \mathcal{L}$\\
$\mathcal{S}$ & Set of SKUs \\
$\mathcal{O}$ & Set of orders \\
$\mathcal{S}(o)$ & Set of SKUs included in order $o \in \mathcal{O}$ \\
$\mathcal{R}_o$ & Set of feasible routes for order $o \in \mathcal{O}$ \\
$\mathcal{F}$ & Set of mandatory assignments of the instance  \\ 
$\mathcal{C}_{os}$ & Set of SL inequalities of order greater than 2 associated with order $o \in \mathcal{O}$ and SKU $s \in \mathcal{S}$\\
\hline
\multicolumn{2}{l}{Parameters} \\ \hline
$K_l$       & Capacity of location $l \in \mathcal{L}$      \\ 
$f(\underline{x^o})$ & Walking to distance to pick all the SKUs of order $o \in \mathcal{O}$ \\
$c_{or}$ & Walking distance of route $r \in \mathcal{R}_o$ servicing order $o \in \mathcal{O}$ \\ 
$b_{or}^l$ & 1 if route $r \in \mathcal{R}_o$ servicing order $o \in \mathcal{O}$ stops in location $l \in \mathcal{L}$, 0 otherwise\\ 
$a_{or}^l$ & Number of picks of route $r \in \mathcal{R}$, servicing order $o \in \mathcal{O}$, in location $l \in \mathcal{L}$\\
$\delta_r(\overline{\mathcal{L}})$ & 1 if route $r \in \mathcal{R}_o$, servicing order $o \in \mathcal{O}$, stops in one of the locations of the subset $\overline{\mathcal{L}} \subset \mathcal{L}$,\\ & 0 otherwise \\
$d_s$ & Demand of SKU $s \in \mathcal{S}$ \\
\hline
\multicolumn{2}{l}{Decision variables}  \\ \hline
$\xi_{ls}$   &  1 if SKU $s \in \mathcal{S}$ is stored in location $l \in \mathcal{L}$, 0 otherwise\\
$x^o_{ij}$ & 1 if the route servicing order $o \in \mathcal{O}$ uses arc $(v_i,v_j) \in \mathcal{E}$, 0 otherwise \\
$u^o_i$ & Auxiliary variable associated with node $v_i \in \mathcal{V}$ for subtour elimination in order $o \in \mathcal{O}$ \\
%\rev{$f^{os}_{ij}$} & \rev{Flow of commodity $s \in \mathcal{S}(o)$ for order $o \in \mathcal{O}$ in the arc $(v_i,v_j) \in \mathcal{E}$} \\
$\rho_{or}$ & 1 if order $o \in \mathcal{O}$ is serviced by route $r \in \mathcal{R}_o$, 0 otherwise \\ 
\hline
\end{tabular}
\end{table}

\subsection{Problem structure and decomposition}\label{sec:decomposition}

% In this section blabla
In this section, we provide a brief polyhedral analysis of the SLAPRP structure in order to motivate the decomposition method presented in this work.

\paragraph{Assignment polytope.}
The first structure that stands out is the block composed of the variables $\xi$, associated with their assignment constraints~\eqref{eq:c1:01}-\eqref{eq:c1:02}. Being structurally close to the assignment polytope, it is not surprising that this substructure possesses integer corner points as well (this result is proven in Appendix~\ref{appendix:assignment_polytope}). In this case, the formulation is already tight, and the so-called \textit{integrality property} establishes that the convexification of this block will not improve the linear relaxation \citep{vanderbeck_reformulation_2010}.

\paragraph{Routing polytopes.}
For each order $o \in \mathcal{O}$, the polytope defined by the variables $x^o$ and the constraints~\eqref{eq:c1:03}-\eqref{eq:c1:06} forms a routing substructure, enforcing that the route defines a valid tour. These blocks have a block-diagonal structure, they are only linked to the assignment variables by the coupling constraints~\eqref{eq:c1:07}. Since the network-flow formulation for the routing structure is known to have a pretty weak linear relaxation, the routing polytopes appear to be appropriate candidates to apply Dantzig-Wolfe (DW) reformulation.

\paragraph{Linking polytope.}
Constraints~\eqref{eq:c1:07} can be seen as a monolithic block of coupling constraints between the routes. Actually, these constraints do not directly link the routes together: for each order $o \in \mathcal{O}$, the routing variables are only linked to the assignment sub-structure. With this observation, it appears that the matrix of constraints~\eqref{eq:c1:07} has a staircase structure, which is amenable to a Benders decomposition \citep{rahmaniani_benders_2017}. %Originally this decomposition method is designed to work with continuous variables in subproblems, in order to use the duality theory to generate feasibility and optimality cuts for the master problem. The approach has also been used when the subproblems are integer programs, but the method becomes less generic and requires more specific design work for the feasibility and optimality cuts \citep{rahmaniani_benders_2017}.

\paragraph{Variable splitting.}
Formulation (C) is defined on a sparse constraint matrix, with both \textit{linking variables} and \textit{coupling constraints}. This particular structure has often been addressed using the Lagrangian relaxation theory, where the decomposition method is called \textit{variable splitting} or \textit{Lagrangian decomposition} \citep{guignard_lagrangean_1987} and relies on the duplication of the linking variables $\xi$, with one duplicate for each order, the equality between the duplicate being dualized in the objective function using Lagrangian multipliers. The recent work of \cite{clausen_consistency_2021} provides a new perspective for this kind of structure, still based on variable splitting but with a DW reformulation of the coupling constraints instead of a Lagrange-based decomposition.

\paragraph{Conclusion.}
In light of this brief analysis, we conclude that the SLAPRP is very structured and prone to decomposition approaches. In the present work, we choose to convexify the routing polytopes and keep the linking constraints in the formulation. The main motivation behind this choice of design is that the subproblems can be modeled as Elementary Shortest Path Problems with Resource Constraints (ESPPRC), a well-studied problem in the literature that has shown its efficiency when embedded in Branch-and-Price algorithms.

\subsection{Extended formulation}

As motivated in the previous section, we convexify the routing polyhedra to take advantage of the block structure of the constraint matrix. For each order $o \in \mathcal{O}$, we define the polyhedron $\overline{X_o}$ as the routing polyhedron for order $o$ (i.e. including constraints \eqref{eq:c1:03}, \eqref{eq:c1:04}, \eqref{eq:c1:05} and \eqref{eq:c1:06} of model (C), and the linear bounds on the routing variables):
\begin{equation}
    \overline{X_o} = \{(x^o_{ij})_{(v_i,v_j) \in \mathcal{E}} \in \{0,1\}^{|\mathcal{E}|}, (u^o_i)_{v_i \in \mathcal{V}} \,|\, \eqref{eq:c1:03},\,\eqref{eq:c1:04},\,\eqref{eq:c1:05},\,\eqref{eq:c1:06},\, \eqref{eq:c1:10}\}
\end{equation}

According to (\cite{desaulniers_column_2005}, Chapter 12) there are two approaches for the DW reformulation, namely convexification and discretization, in this Section we use the latter. Since $\overline{X_o}$ is bounded, it contains a finite number of points $(\overline{x^{or}})_{r \in \mathcal{R}_o}$, indexed by $r \in \mathcal{R}_o$. With the application of the theorem [\cite{vanderbeck_reformulation_2010} 13.2], inspired by the Minkowski-Weyl theorem, we can write the set $\overline{X_o}$ as:
\begin{equation}\label{eq:dw_reformulation:01}
    \overline{X_o} = \{x \in \mathbb{R}^{|\mathcal{E}|} \;|\; x = \sum_{r \in \mathcal{R}_o} \rho_{or}\overline{x^{or}},\, \sum_{r \in \mathcal{R}_o} \rho_{or} = 1,\, \rho_{or} \in \{0,1\}, \,\forall r \in \mathcal{R}_o\}
\end{equation}

We can then optimize over the convex hull of these polyhedra, instead of their linear relaxation. This is done by substituting the $x^o$ variables of formulation (C) by their expression in equation~\eqref{eq:dw_reformulation:01}. Let us introduce $c_{or} = f(\overline{x^{or}})$ the walking distance of route $r \in \mathcal{R}_o$ and $a_{or}^l = \sum_{v_i \in \mathcal{V}^0} \sum_{v_j \in \mathcal{V}(l)} \overline{x^{or}_{ij}}$, that is the number of picks at location $l \in \mathcal{L}$ for route $r$. The reformulation leads to formulation (DW), which we refer to as the DW reformulation of (C):
\begin{align}
    & \textbf{(DW)} &\min\, & \sum_{o \in \mathcal{O}}\sum_{r \in \mathcal{R}_o}  c_{or}\rho_{or} & &\label{eq:dw2:obj} \\
        & & s.t.\, & \sum_{s \in \mathcal{S}} \xi_{ls} \leq K_l   & \forall l \in \mathcal{L} & \label{eq:dw2:01} \\
        & & & \sum_{l \in \mathcal{L}} \xi_{ls} = 1  & \forall s \in \mathcal{S} & \label{eq:dw2:02}\\
        & & & \xi_{ls} = 1 & \forall (s,l) \in \mathcal{F} & \label{eq:dw2:022}\\
        & & & \sum_{r \in \mathcal{R}_o} \rho_{or} = 1 & \forall o \in \mathcal{O} & \label{eq:dw2:03}\\
        & & & \sum_{r \in \mathcal{R}_o} a_{or}^l\rho_{or} \geq \sum_{s \in \mathcal{S}(o)} \xi_{ls} & \forall l \in \mathcal{L}, o \in \mathcal{O} & \label{eq:dw2:04} \\
        & & & \rho_{or} \in \{0,1\} & \forall o \in \mathcal{O}, r \in \mathcal{R}_o & \label{eq:dw2:05}\\
        & & & \xi_{ls} \in \{0,1\} & \forall l \in \mathcal{L}, s \in \mathcal{S}&\label{eq:dw2:06}
\end{align}
The objective function~\eqref{eq:dw2:obj} is the sum of the distances of the selected routes in the solution. Note that, compared to formulation (C), the objective function becomes linear without any additional assumption. Constraints~\eqref{eq:dw2:01} and~\eqref{eq:dw2:02} are the assignment constraints and ensure that the capacity of each location is satisfied and each SKU is placed in one location. Constraints~\eqref{eq:dw2:03} ensure that exactly one route is chosen to retrieve each order. Constraints~\eqref{eq:dw2:04} ensure that the route that collects $o \in \mathcal{O}$ stops at location $l \in \mathcal{L}$ if there is an SKU $s \in \mathcal{S}(o)$. They also ensure that the number of stops in this location is consistent with the number of SKUs of the order in this location.

The LP relaxation of (DW) is obtained by replacing binary requirements with non-negativity requirements on variables $\rho$ and $\xi$. It is referred to as the \textit{Master Problem} (MP). When the MP is defined over subsets of routes $\overline{\mathcal{R}_o}\subset\mathcal{R}_o$ for at least one order $o\in\mathcal{O}$ we call it the \textit{Restricted Master Problem} (RMP).

\section{Strengthened Linking Inequalities}\label{sec:og_cuts}

In this section, we introduce a family of valid inequalities for (DW) that tighten its linear relaxation by strengthening the link between assignment variables and routing variables.

\begin{theorem}\label{prop:sl}
Given $\overline{\mathcal{L}}\subset\mathcal{L}$, the following inequalities (coined SL inequalities) are valid for formulation (DW)
\begin{equation}\label{eq:el_cut}
        \sum_{r \in \mathcal{R}_o} \delta_r(\overline{\mathcal{L}}) \rho_{or} \geq \sum_{l \in \overline{\mathcal{L}}} \xi_{ls} \qquad \forall o \in \mathcal{O}, s \in \mathcal{S}(o)
    \end{equation}
    where $\delta_r(\overline{\mathcal{L}})$ equals 1 if route $r \in \mathcal{R}_o$, $o \in \mathcal{O}$, stops in one of the locations in $\overline{\mathcal{L}} \subset \mathcal{L}$, 0 otherwise.
\end{theorem}

\paragraph{Proof.} See Appendix~\ref{appendix:sl_proof}.

Note that the SL inequalities are exponential in number and non-robust in the general case (i.e., when $|\overline{\mathcal{L}}| \geq 2$).
In the following, we will refer to the number of elements in the set $\overline{\mathcal{L}}$ as the \textit{order} of the inequality. We note $\mathcal{C}_{os}$ the set of SL inequalities of order greater than or equal to 2 associated with the order $o \in \mathcal{O}$ and the SKU $s \in \mathcal{S}(o)$. The special case when the order of the SL is equal to 1 is studied in the next section.

% special case of order 1

\subsection{Special case SL inequalities of order 1.}\label{sec:ineq:sl1}
In the general case, the SL inequalities are non-robust, and their number is exponential. However, SL inequalities of order 1 (SL-1, in the following) are robust with the proposed formulation, and their number is polynomial. Let us define the parameter $b_{or}^l$ that equals 1 if route $r \in \mathcal{R}_o$ stops at location $l \in \mathcal{L}$, 0 otherwise. The SL-1 family of inequalities can be expressed as follows:
\begin{equation}\label{eq:order1_el}
    \sum_{r \in \mathcal{R}_o} b_{or}^l\rho_{or} \geq \xi_{ls} \quad \forall l \in \mathcal{L}, o \in \mathcal{O}, s \in \mathcal{S}(o)
\end{equation}

As constraints~\eqref{eq:dw2:04} of (DW), inequalities~\eqref{eq:order1_el} link assignment and routing variables. The difference lies in the way stops are counted: a route that stops twice in a location is counted once on the left-hand side of inequalities~\eqref{eq:order1_el}, while it is counted twice in the constraints~\eqref{eq:dw2:04}. 

Note that the addition of SL-1 is crucial in the design of an efficient branching scheme (details will be provided in Section~\ref{sec:branching}). Indeed the following theorem holds.

\begin{theorem}\label{prop:branching}
Let $X = (\xi,\rho)$ be an optimal solution of the master problem, strengthened by SL-1 inequalities. If variables $\xi$ are integral, then $X$ is integral, or there exists an integral solution $X'$ of the same cost.
\end{theorem}

\paragraph{Proof.} See Appendix~\ref{appendix:branching_proof}.

\begin{figure}[h]
\centering
    \begin{minipage}{.45\textwidth}
    \centering
    \includegraphics[scale=1.2]{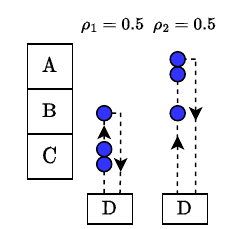}
    \end{minipage}
    \begin{minipage}{.45\textwidth}
    \centering
    \includegraphics[scale=1.2]{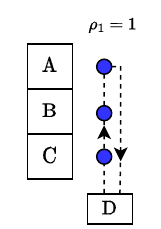}
    \end{minipage}
    \caption{Effect of SL-1 inequalities on branching decisions}
    \label{fig:EL_illustration}
\end{figure}

Theorem~\ref{prop:branching} is illustrated by Figure~\ref{fig:EL_illustration}. One order is composed of three SKUs (A, B, C), assigned to the three locations (i.e., $\xi$ variables are integral). On the left, without SL-1 inequalities, the solution is not integer as two routes are used to pick the order, each with $\rho = 0.5$, and a total distance of $0.5 \cdot 4 + 0.5 \cdot 6 = 5$. On the right, with SL-1 inequalities, the solution uses a single route, for a distance of 6.

Since SL-1 inequalities are of polynomial size and robust, they are all added to the master problem formulation at the beginning of the resolution. 

% separation
\subsection{Separation of SL inequalities}

The SL inequalities are embedded in a cutting plane framework, where only violated cuts are added to cut off fractional solutions. Note that we can decompose the SL separation problem $(SL-SEP)$ in subproblems $(SL-SEP)_{os}$ defined for each order $o \in \mathcal{O}$ and SKU $s \in \mathcal{S}(o)$. Each $(SL-SEP)_{os}$ consists in finding a subset of storage locations $\overline{\mathcal{L}} \subset \mathcal{L}$ that leads to the most violated (if any) SL inequality. Given an optimal solution $(\xi^*_{ls}, \rho^*_{or})$ of the RMP, and using the binary variables $x_l$ to model whether $l \in \overline{\mathcal{L}}$, and $y_r$ to model whether route $r \in \mathcal{R}_o$ stops at one of the chosen locations $\overline{\mathcal{L}}$, $(SL-SEP)_{os}$, $o \in \mathcal{O}$, $s \in \mathcal{S}(o)$ can be stated as the following binary integer program:
\begin{align}
    & \textbf{$(SL-SEP)_{os}$} &Max \; & \sum_{l \in \mathcal{L}} \xi^*_{ls}x_l - \sum_{r \in \mathcal{R}_o} \rho^*_{or}y_r &  &\\
    & &s.t.\; & a_{or}^lx_l \leq y_r & \forall r \in \mathcal{R}_o, l \in \mathcal{L} &\label{eq:slsep_01}\\
    & & & x_l \in \{0,1\} & \forall l \in \mathcal{L} & \label{eq:slsep02} \\
    & & & y_r \in \{0,1\} & \forall r \in \mathcal{R}_o & \label{eq:slsep03}
\end{align}

If the optimal solution of $(SL-SEP)_{os}$ has a strictly positive value, then there exists an SL inequality that cuts the solution $(\xi^*,\rho^*)$. If the optimal solution of $(SL-SEP)_{os}$ is negative for all $o \in \mathcal{O}, s \in \mathcal{S}(o)$, then there exists no such cut. 

$(SL-SEP)_{os}$ is solved by dynamic programming as follows. Let us define $\mathcal{P}_l(\widetilde{\mathcal{R}_o})$ the optimal value of $(SL-SEP)_{os}$ restricted to the subset of locations $\{1,\dots,l\} \subset \mathcal{L}$ and the subset of routes $\widetilde{\mathcal{R}_o} \subset \mathcal{R}_o$. The value of $\mathcal{P}_l(\widetilde{\mathcal{R}_o})$ can be computed according to the following equation:

\begin{equation}\label{eq:slsep_recu}
    \mathcal{P}_l(\widetilde{\mathcal{R}_o}) = \max \{ \; \mathcal{P}_{l-1}(\widetilde{\mathcal{R}_o}) : \mathcal{P}_{l-1}(\{ r \in \widetilde{\mathcal{R}_o} \, |\, a_{or}^l = 0\}) + \xi_{ls}^* - \sum_{r \in \widetilde{\mathcal{R}_o}} a_{or}^l \rho_{or}^* \;\} 
\end{equation}
\noindent
with the initialization $\mathcal{P}_0(\widetilde{\mathcal{R}_o}) = 0$. At each step of the recurrence, a binary choice takes place:
\begin{itemize}
    \item The location $l$ is not added to the set $\overline{\mathcal{L}}$. In this case, the optimal value over locations $\{1, \dots, l\}$ is equal to the optimal value over $\{1, \dots, l-1\}$, i.e., $\mathcal{P}_{l-1}(\widetilde{\mathcal{R}_o})$.
    \item The location $l$ is added to the set $\overline{\mathcal{L}}$. In this case, $\mathcal{P}_l(\widetilde{\mathcal{R}_o})$ is obtained from $\mathcal{P}_{l-1}(\widetilde{\mathcal{R}_o})$ by adding $\xi_{ls}^*$ minus $\rho_{or}^*$ for each route $r\in\widetilde{\mathcal{R}_o}$  that stops at $l$, and removing said routes from $\widetilde{\mathcal{R}_o}$.
\end{itemize}

\paragraph{Implementation details.} 
The dynamic programming algorithm is solved for each $o \in \mathcal{O}$ and $s \in \mathcal{S}$ and in order not to spend a prohibitive amount of time on this procedure, we enhance it as follows. First, the list of locations $\mathcal{L}$ is sorted in increasing order of $\xi_{ls}^*$ so that the locations with the highest potential of generating a violated cut are explored first. This helps to prune the search (see next point).  Second, we prune the search in three cases. In the first case, we check if the sum of the remaining $\xi^*$ is too low to get a violated cut. In the second case, if all the remaining $\xi^*$ are equal to zero and the current cut is already violated, we stop the search since it is not possible to get a deeper cut. In the third case, if all remaining $\rho^*$ are equal to zero, the objective value is increased by the sum of the strictly positive remaining $\xi^*$.

\section{Branch-Cut-and-Price algorithm}\label{sec:algorithm}

A Branch-Cut-and-Price algorithm (BCP) is a Branch-and-Bound algorithm in which the dual bound of each node of the tree is computed using the linear relaxation of an extended formulation, solved by column generation, and strengthened by the addition of cutting planes \citep{desaulniers_column_2005}. In this section, we describe the main components of the algorithm, namely, the column and cut generation procedure (see Section~\ref{sec:cg}), the cut management scheme (see Section~\ref{sec:cut_management}), the primal heuristic used at each node of the Branch-and-Bound tree (see Section~\ref{sec:primal}), the branching scheme (see Section~\ref{sec:branching}), and the strengthening procedure for branching constraints (see Section~\ref{sec:symmetry}).

\subsection{Column and cut generation}\label{sec:cg}

For each order $o \in \mathcal{O}$, let $\overline{\mathcal{R}_o} \subset \mathcal{R}_o$ be a subset of the feasible routes for $o$, and we note $\overline{\mathcal{R}} = (\overline{\mathcal{R}_1}, \overline{\mathcal{R}_2}, \dots , \overline{\mathcal{R}_{|\mathcal{O}|}})$ the collection of these subsets. Let $\overline{\mathcal{C}_{os}} \subset \mathcal{C}_{os}$ be a subset of the SL inequalities of order greater than or equal to 2 associated with $o \in \mathcal{O}$ and $s \in \mathcal{S}(o)$, and we note $\overline{\mathcal{C}} = (\overline{\mathcal{C}_{os}})_{o \in \mathcal{O},s \in \mathcal{S}(o)}$. The RMP defined on $\overline{\mathcal{R}}$, and strengthened by the valid inequalities SL-1 and $\overline{\mathcal{C}}$, is noted $RMP(\overline{\mathcal{R}}, \overline{\mathcal{C}})$.
\begin{align}
    & \textbf{$(RMP(\overline{\mathcal{R}}, \overline{\mathcal{C}}))$} &\min\, & \sum_{o \in \mathcal{O}}\sum_{r \in \overline{\mathcal{R}_o}}  c_{or}\rho_{or} & & \\
        & &s.t.\, & \sum_{s \in \mathcal{S}} \xi_{ls} \leq K_l   & \forall l \in \mathcal{L} & \\
        & & & \sum_{l \in \mathcal{L}} \xi_{ls} = 1  & \forall s \in \mathcal{S} & \label{eq:rmp:05} \\
        & & & \xi_{ls} = 1 & \forall (s,l) \in \mathcal{F} & \\
        & & & \sum_{r \in \overline{\mathcal{R}_o}} \rho_{or} = 1 & \forall o \in \mathcal{O} & \quad (\mu_o) \label{eq:rmp:01} \\
        & & & \sum_{r \in \overline{\mathcal{R}_o}} a_{or}^l\rho_{or} \geq \sum_{s \in \mathcal{S}(o)} \xi_{ls} & \forall l \in \mathcal{L}, o \in \mathcal{O} & \quad (\pi_{lo}) \label{eq:rmp:02} \\
        & & & \sum_{r \in \overline{\mathcal{R}_o}} b_{or}^l\rho_{or} \geq \xi_{ls} & \forall l \in \mathcal{L}, o \in \mathcal{O}, s \in \mathcal{S}(o) & \quad (\sigma_{osl}) \label{eq:rmp:03} \\
        & & & \sum_{r \in \overline{\mathcal{R}_o}} \delta_r(\mathcal{L}_c) \rho_{or} \geq \sum_{l \in \mathcal{L}_c} \xi_{ls} & \forall o \in \mathcal{O}, s \in \mathcal{S}(o), c \in \overline{\mathcal{C}_{os}} & \quad (\lambda_{osc}) \label{eq:rmp:04}\\
        & & & \rho_{or} \geq 0 & \forall o \in \mathcal{O}, r \in \overline{\mathcal{R}_o} & \\
        & & & \xi_{ls} \geq 0 & \forall l \in \mathcal{L}, s \in \mathcal{S}&
\end{align}

Constraints~\eqref{eq:rmp:03} correspond to SL-1 inequalities, that are all added to the master formulation from the start, and Constraints~\eqref{eq:rmp:04} correspond to the active SL inequalities of order greater than or equal to 2 at the current iteration. At each iteration, the pricing problems find columns with a negative reduced cost that are then added to $(RMP(\overline{\mathcal{R}},\overline{\mathcal{C}}))$, or prove that none exists. Dual variables associated with Constraints~\eqref{eq:rmp:01}-~\eqref{eq:rmp:04} are given in parentheses. The reduced cost of the route $r \in \mathcal{R}_o$ is defined as follows:

\begin{equation}
    \overline{c_{or}} = c_{or} - \mu_o - \left( \sum_{l \in \mathcal{L}} a_{or}^l\pi_{lo} \right) 
    - \left( \sum_{l \in \mathcal{L}} b_{or}^l \left( \sum_{s \in \mathcal{S}(o)} \sigma_{osl} \right) \right)
    - \left( \sum_{c \in \overline{\mathcal{C}}} \delta_r(\mathcal{L}_c) \left(\sum_{s \in \mathcal{S}(o)} \lambda_{osc} \right) \right)
\end{equation}

The algorithm used to solve the pricing problem is presented in Section~\ref{sec:pricing}. When there are no more columns with negative reduced costs, the separation problems are solved to find violated SL inequalities (see Section~\ref{sec:og_cuts}). 

\paragraph{Management of infeasible linear programs.}
Since the SLAPRP is mostly free from operational constraints, feasibility is generally not an issue. However, it might happen that at a given node of the Branch-and-Bound tree, the current set of columns does not allow to obtain a feasible solution for the restricted master problem $RMP(\overline{\mathcal{R}},\overline{\mathcal{C}})$. This may happen when the branching constraints cut off all feasible columns for a given order. To prevent this situation, we introduce a set of \textit{super columns} that ensure that the $RMP(\overline{\mathcal{R}},\overline{\mathcal{C}})$ remains feasible at each node of the Branch-and-Bound tree. For each order $o \in \mathcal{O}$, a super column $r_o^{SUP}$ is defined as a route stopping in every storage node in $\mathcal{G}$, i.e., $a_l^{r_o^{SUP}} = K_l$ for all $l \in \mathcal{L}$. In this case, the solution $\rho_{or_o^{SUP}} = 1$ for each order $o \in \mathcal{O}$ is a feasible solution of $RMP(\overline{\mathcal{R}},\overline{\mathcal{C}}))$, no matter the value of the assignment variables. However, these super routes are obviously not valid, violating Constraints~\eqref{eq:c1:06} of the compact formulation. The cost of these routes is thus set to a big enough value, so that they are not part of any optimal solution of the $RMP(\overline{\mathcal{R}},\overline{\mathcal{C}}))$, after convergence of the column generation. 

\subsection{Management of the SL cuts}\label{sec:cut_management}

The SL inequalities are non-robust with the formulation of the master problem, i.e., they cannot be expressed with the network-flow variables of the compact formulation (C). In consequence, the addition of one SL cut modifies the structure of the pricing problem: As explained in Section~\ref{sec:pricing}, it is modeled as an ESPPRC and it will need an additional resource to account for the cut, weakening the dominance criterion. To keep the size of the formulation reasonable, and the pricing problems tractable, careful management of the SL cuts appears as a necessity.

\paragraph{Cut pool.} SL inequalities are added dynamically to the RMP formulation via a cut pool. Whenever an SL inequality is separated, it is kept in the cut pool. Every time $RMP(\overline{\mathcal{R}},\overline{\mathcal{C}})$ is solved, the cut pool is checked for violated inequalities. If a violated inequality is found, it is added to the set of active cuts $\overline{\mathcal{C}}$, and $RMP(\overline{\mathcal{R}},\overline{\mathcal{C}})$ is solved again. At the same time, a procedure determines which of the active SL inequalities in $\overline{\mathcal{C}}$ are satisfied at equality in the current solution. If an inequality is not binding, it is removed from the formulation. If the cut pool contains no violated inequality, the separation procedure is called. If violated cuts are found, they are added to the cut pool. Then, some of them, depending on the number of separated inequalities (see next paragraph), are added to the set of active cuts $\overline{\mathcal{C}}$, and the RMP is solved again.

\paragraph{Implementation details.} On top of the cut pool management, several related features in our implementation have been crucial in the performance of our algorithm. First, the separation procedure is only called during the resolution of nodes up to a depth of 3 in the Branch-and-Bound tree. A problematic behavior we identified in preliminary experiments is that, for some instances, the separation of the SL inequalities returns a prohibitively large number of violated cuts. If they are all added to the formulation, it becomes too large to handle, making the master or pricing problem intractable. To tackle this issue, we limit at 500 the number of simultaneously active SL cuts for one order. If too many of them are separated, only the most violated are added to the formulation. Experiments show that the feature does not impact much the bound, as the number of active cuts decreases to a manageable amount with the convergence of the column generation.

\subsection{Primal heuristics}\label{sec:primal}

At each node of the Branch-and-Bound tree, a primal heuristic procedure is called in order to improve the primal bound of the tree. Indeed, integer solutions of $RMP(\overline{\mathcal{R}},\overline{\mathcal{C}})$ are typically encountered deep in the Branch-and-Bound tree. If the starting primal bound is of poor quality, it would remain unchanged for a large part of the running time and prevent an efficient pruning of the tree. To mitigate this effect, a basic constructive heuristic is run at each node of the tree. Since the problem is loosely constrained, a simple heuristic with a low computational burden proved to be sufficient for this purpose. At a given node of the tree, we first solve the column and row generation procedure, then we proceed as follows. 
We start from a partial solution that consists of the fixed locations $\mathcal{F}$ of the instance and those imposed by the branching decisions at the current node. Then we insert SKUs one by one by selecting, at each step, the pair $(l,s)$,  $l \in \mathcal{L}$, $s \in \mathcal{S}$ that constitutes a feasible update and maximizes the value of $\xi_{ls}$ in the optimal solution of the RMP at the current node of the Branch-and-Bound tree. In other terms, we assign the SKU associated with the least fractional $\xi$ variable at each iteration. The procedure stops when all SKUs have been inserted. 

\paragraph{Initial set of columns.}
At first, a heuristic solution is calculated with the aim of obtaining an initial set of columns.
%The column generation procedure is started with a heuristic solution, providing the initial set of columns in the pool. Since a primal heuristic is launched at each node of the tree, the quality of the incumbent solution has little impact on the performances of the algorithm. Moreover, 
Note that any assignment that enforces the fixed positions of the instance leads to a feasible solution. Thus, the starting solution is the result of a basic Hill-Climbing heuristic, where a random (feasible) solution is improved at each iteration by the exchange of two SKUs. If no such improvement exists, the procedure is stopped.

\subsection{Branching scheme}\label{sec:branching}

According to Theorem~\ref{prop:branching}, the integrality of the $\xi$ variables guarantees the integrality of the solution, we therefore only need to branch on such variables. Another advantage of branching on the $\xi$ variables is that the pricing problems are unaffected by the branching decisions. Based on these observations, a natural branching scheme, called \textit{location branching}, would be to branch on the most fractional $\xi$ variable. However, preliminary experiments show that this strategy produces unbalanced search trees. To overcome this drawback, we propose a two-level branching strategy, called \textit{combined branching}. At the first level, we branch on the assignment of SKUs in entire aisles (instead of single locations). Since integer aisle locations are not sufficient to restore integrality in the solutions, we use the location branching at the second level. Numerical experiments in Section~\ref{sec:experiments} provide a comparison between the location and combined strategies.

\paragraph{Location branching.}
In this scheme, we branch on a $\xi$ variable. Let $(\xi,\rho)$ be a solution of $RMP(\overline{\mathcal{R}}, \overline{\mathcal{C}})$. If this solution is not integer, there exists at least one fractional variable $\xi_{ls}$ with $l \in \mathcal{L}$ and $s \in \mathcal{S}$. 
The variable $\xi_{ls}$ that maximizes the quantity
$$ d_{s} \cdot \min(\xi_{ls} \; , \; (1 - \xi_{ls})),$$
is selected to perform the branching. Here $d_s = | \{o \in \mathcal{O} \, | \, s \in \mathcal{S}(o)\}|$ is the demand of $s$. This means that we choose the most fractional variable weighted by the demand of the associated SKU. Indeed, the $\xi$ variables corresponding to an SKU with a large demand will appear in more orders, and then in more constraints. Fixing its value is more likely to have a high impact on the solution in children nodes. We derive two branches fixing the variable to one (meaning $s$ is assigned to the location $l$), and fixing it to zero (the assignment is forbidden). Equation~\eqref{eq:branching_disjunction1} provides the corresponding disjunction we impose in the children nodes. 
\begin{equation}\label{eq:branching_disjunction1}
    (\xi_{ls} \geq 1) \lor (\xi_{ls} \leq 0)
\end{equation}

\paragraph{Combined branching.}
Let us index by $\{1,\dots,\overline{a}\}$ the different aisles of the layout and let us indicate by $\mathcal{L}^a$ the set of locations in aisle $a=1,\dots,\overline{a}$. For an SKU $s \in \mathcal{S}$, and an aisle $1 \leq a \leq \overline{a}$, we propose to branch on whether $s$ will be assigned to a location in aisle $a$, or not, that is we branch on the following disjunction:
$$\left(\sum_{l \in \mathcal{L}^a} \xi_{ls} \leq 0\right) \lor \left(\sum_{l \in \mathcal{L}^a} \xi_{ls} \geq 1\right).$$
A heuristic picks the candidate aisle $a = 1,\dots,\overline{a}$ and the candidate SKU $s \in \mathcal{S}$ maximizing the following quantity:
$$ d_{s} \cdot \min(\sum_{l \in \mathcal{L}^a}\xi_{ls} \; , \; (1 - \sum_{l \in \mathcal{L}^a}\xi_{ls}))$$
Branching on a single location leads to an unbalanced tree, with the branch where a variable is set to zero being quite weak. This is especially true for the SLAPRP variants where aisles are crossed entirely (i.e., S-shape, midpoint, and largest gap policies): forbidding the assignment of an SKU to a location might lead to a symmetric solution obtained by switching the partial assignment within an aisle. Thus forbidding the assignment of an SKU to an entire aisle may have a larger impact on the solution, and leads to tighter bounds in the children nodes. 

Note, however, that the aisle branching is not sufficient to ensure integrality, as a solution can have all SKUs fixed to aisles without being integral. To address this drawback, we use the location branching scheme at the second level of the combined strategy. If the best candidate with aisle branching is associated with a quantity that is below a threshold of 0.25, it means that the SKUs are all close to being fixed on aisles, and we use the location branching scheme instead.

\paragraph{Implementation details.}
In all instances used for experiments, the layout is a rectangular warehouse, with integer distances. If we note $D^a$ the distance between two consecutive aisles and $D^b$ the distance between two consecutive locations, it appears that an integer solution can only have an objective value that is a multiple of $2 \cdot gcd(D^a,\,D^b)$ (the proof of this result is straightforward). Therefore, we can round up any valid dual bound accordingly, ensuring a stronger pruning routine and a better gap.

\subsection{Strengthening procedure for branching constraints}\label{sec:symmetry}

Some instances are characterized by a high degree of symmetry, which may lessen the impact of the branching schemes. In the following, we introduce a procedure to strengthen the branching decisions by breaking symmetries. This method is inspired by the work of \cite{margot_pruning_2002} on isomorphic pruning, and the orbital branching scheme introduced by \cite{ostrowski_orbital_2011}. The strengthening procedure is a simplified version of these works, applied to a straightforward symmetry. We describe the scheme on a branching disjunction on a single variable (location branching), but the same reasoning remains valid with aisle branching.

If two SKUs $s^1, s^2 \in \mathcal{S}$ appear in the same orders, i.e., $\mathcal{O}(s^1) = \mathcal{O}(s^2)$, we say they are \textit{symmetric}. This means that for a valid solution, either integer or fractional, permuting the values of the variables $\xi_{ls^1}$ and $\xi_{ls^2}$ for all $l \in \mathcal{L}$ leads to another valid solution with the same objective value. This result is straightforward since all routes that need to pick one SKU also need to pick the other. Thus permuting their locations will not change the chosen routes.

We note $F^1$ the set of variables that have already been branched on, and set to 1, in the current node of the Branch-and-Bound tree, and $sym(s)$ the set of SKUs symmetric with $s$ (including $s$). Let us consider $\{s^1,\dots, s^k \} = sym(s)\backslash F^1$ the set of SKUs that are symmetrical with $s$, and have not been fixed to one in the current node. Instead of enforcing the disjunction~\eqref{eq:branching_disjunction1} introduced in the variable branching scheme, we consider the disjunction~\eqref{eq:branching_disjunction2} that gives a partition of the integer search space where either one of the SKUs of $sym(s)$ is assigned to $l$, or all these assignments are forbidden.
\begin{equation}\label{eq:branching_disjunction2}
    (\xi_{ls^1} \geq 1) \lor \dots \lor (\xi_{ls^k} \geq 1) \lor (\sum_{1 \leq i \leq k} \xi_{ls^i} \leq 0).
\end{equation}
This disjunction contains more than the usual two branches. However, since the SKUs $s^1, \dots, s^k$ are symmetric, the different branches $(\xi_{ls^i} \geq 1)$ for $1 \leq i \leq k$ lead to equivalent problems. In this case, we can only consider one of those branches, e.g. $(\xi_{ls} \geq 1)$, and discard the others without loss of generality. The strengthened branching scheme then enforces the following disjunction:
\begin{equation}\label{eq:branching_disjunction3}
    (\xi_{ls} \geq 1) \lor (\sum_{1 \leq i \leq k} \xi_{ls^i} \leq 0).
\end{equation}

\section{Solution approach for the pricing problem}\label{sec:pricing}

The pricing problems for the SLAPRP are modeled as Elementary Shortest Path Problems with Resource Constraints (ESPPRC), which is a popular design in the column generation literature \citep{desaulniers_column_2005}. Note that we convexified the operational details of the problem to obtain a generic formulation, as described in Section~\ref{sec:description_and_formulation}, and as a consequence, they appear in the subproblems. The backbone of the algorithm is the same no matter the configuration. The differences appear in:
\begin{itemize}
    \item the underlying graph $\mathcal{G}$ that models the warehouse layout;
    \item the resource extension function, and, if needed, additional parameters for the routing policy.
\end{itemize}

This section is organized as follows: We first introduce the problem in Section~\ref{sec:pricing:description}. In Section~\ref{sec:pricing:labeldef}, we introduce our algorithm and the label definition for the \textit{optimal routing policy}. The label extension function is described in Section~\ref{sec:label_extension}, and the dominance rules in Section~\ref{sec:dominance}. Then, we highlight the differences induced by other routing policies in Section~\ref{sec:pricing_other_policies}.

\subsection{Problem description}\label{sec:pricing:description}

For an order $o \in \mathcal{O}$, the $o^{th}$ subproblem consists in finding routes of negative reduced cost, as introduced in Section~\ref{sec:cg}, or proving that no such route exists. The pricing problem is defined on the graph $\mathcal{G}$ introduced in Section~\ref{sec:problem_description}. A valid solution consists in an elementary tour on graph $\mathcal{G}$, which includes exactly $|\mathcal{S}(o)| + 1$ edges, and stops in the mandatory locations corresponding to the SKUs of $o$ that are assigned to fixed positions. 
Furthermore, we impose what we call the \textbf{first stop restriction}, which enforces a route to stop at a location the first time it passes in front of it. Due to the fact that the distance matrix satisfies the triangle inequality, it is clear that all the routes in any optimal solution satisfy the condition. Note that its enforcement is optional in the problem definition, but it limits the propagation of labels by exploiting the Steiner graph structure of the warehouse layout to discard symmetric labels.

A notable difference between the pricing problem of the SLAPRP and classical ESPPRCs for vehicle routing problems lies in the resource constraints. They are all expressed as \textit{equality constraints}, instead of the more classical capacity or time window constraints: For a path to be valid, the resource corresponding to its length (i.e., the number of edges traversed by the path) must be equal to $|\mathcal{S}(o)| + 1$ at the sink node, and the resources corresponding to the visit of mandatory locations must be consumed. This feature is unusual but does not impact much the solution approach. Indeed, \cite{irnich_resource_2007} highlights that these constraints can be modeled with classical resource extension functions, and all the theory on the ESPPRC remains valid. The pricing problems are solved using a labeling algorithm, where labels represent partial routes extended from the source node (the drop-off point) to the sink node (copy of the drop-off point). A label is extended from a node to its successors using resource extension functions, and a dominance criterion is used to discard labels that cannot lead to optimal solutions. The active SL cuts being non-robust, each of them is managed as a single resource: one active cut is characterized by a set of locations, and the associated dual cost is counted if at least one of these locations is visited by the path. The other constraints being robust, their dual costs simply add additional terms in the distance matrix of graph $\mathcal{G}$.

\subsection{Label definition for optimal routing}\label{sec:pricing:labeldef}

A label $L$ represents a path $P$ starting at the drop-off point in graph $\mathcal{G}$. It is defined by a vector $L = (v,c,q,R,T,F)$, where $v$ is the last node visited in $P$, $c$ is the current reduced cost of $P$, $q$ is the number of edges included in $P$, $R$ is the location reachability vector, $T$ is the cut reachability vector, and $F$ is the mandatory location vector. Resource $R$ is defined with three possible states for each location $l \in \mathcal{L}$: (i) $R_l = V$ if location $l$ has been visited by the path, (ii) $R_l = R$ if location $l$ is reachable in the current path, and (iii) $R_l = U$ if location $l$ is unreachable in the current path. Note that, apart from already visited locations, all the other locations are still technically reachable, but the first stop restriction allows us to derive a stronger reachability condition. Resource $T$ is similarly defined with three possible states for each active SL cut $c \in \overline{\mathcal{C}}$: (i) $T_c = V$ if the corresponding dual cost is already counted in the reduced cost of $P$, (ii) $T_c = R$ if the dual cost is still potentially reachable for the path $P$, and (iii) $T_c = U$ if it is not reachable anymore. Since the SL cuts are non-robust, we need to keep in memory which cuts have been counted in the current path, in order not to account for more than once their contribution to the reduced cost. In order to get stronger dominance rules, we also check if a cut is not reachable anymore. Resource $F$ is defined as a vector where each location is associated with the corresponding number of remaining mandatory stops. A valid path consumes the resources in $F$ completely, i.e., the mandatory locations should all be visited, potentially several times if required.

\subsection{Label extension for optimal routing}\label{sec:label_extension}

In this section, we describe the label extension function in the case of optimal routing. Let $L^1 = (v^1,c^1,q^1,R^1,T^1,F^1)$ be a label attached to node $v^1 \in \mathcal{V}^0$, that is extended to node $v^2 \in \mathcal{V}$. We note $l^1,l^2 \in \mathcal{L}$ the locations associated with the nodes $v^1$ and $v^2$. First, the extension is possible if and only if it satisfies the following conditions: (i) $q^1 < |\mathcal{S}(o)|$ since labels are only extended up to a length of $|\mathcal{S}(o)|$. (ii) $(v^1,v^2) \in \mathcal{E}$, i.e. there exists an arc from $v_i$ to $v_j$. (iii) $v^2$ is reachable in the current label according to the location reachability vector, i.e. $R^1_{l^2} = R$. (iv) $F^1_{l^2} > 0$ or $q^1 + \sum_{l \in \mathcal{L}}F^1_l < |\mathcal{S}(o)|$. This condition ensures that $L^1$ is only extended if $l^2$ is a mandatory location, or if the length of $L^1$ allows one to visit all the remaining mandatory stops after visiting $l^2$. (v) The extension to $v^2$ will not lead to an unfeasible route considering the fixed assignments. In other terms, let us consider that there are already $k > 0$ SKUs assigned to the location $l^2$ (either by branching decisions or fixed assignments on the instance) that are not part of $o$. Then the route cannot stop more than $K_l - k$ times in $l$.

If $v^2$ satisfies these conditions, then the extension is possible, and the new label $L^2$ is defined as $L^2 = (v^2,c^2,q^2,R^2,T^2,F^2)$ such that: (i) $R^2 \gets R^1$. (ii) $\forall l \in \mathcal{L}$: if $l = l^2$, then $R^2_l \gets V$, else if $l$ is located on the shortest path from $l^1$ to $l^2$, then $R^2_l \gets U$. The second condition corresponds to the first stop restriction presented in Section~\ref{sec:pricing:description}. (iii) $T^2 \gets T^1$. (iv) $\forall c \in \overline{\mathcal{C}}$: If $T^1_c = U$, then $T^2_c \gets U$. Else if $l^2 \in \mathcal{L}_c$ (i.e., label $L^2$ gets the dual value associated to cut $c$), then $T^2_c \gets V$. Else if $R^2_l = U, \, \forall l \in \mathcal{L}_c$, (i.e. all locations that define cut $c$ are either visited or unreachable in the current path), then $T^2_c \gets U$. (v) $F^2 \gets F^1$. (vi) If $F^2_{l^2} > 0$ then $F^2_{l^2} \gets F^2_{l^2} - 1$. (vii) $c^2 \gets c^1 + c(v^1,v^2) + \pi_{l^2o} + \mathbb{1}_{l^1 \neq l^2}\sum_{s \in \mathcal{S}(o)}\delta_{osl^2} + \sum_{c \in \mathcal{C}^*} \lambda_c$ with $\mathbb{1}$ being an indicator parameter, and $\mathcal{C}^* = \{c \in \overline{\mathcal{C}} \;|\; T^1_c = R \, \& \, T^2_c = V\}$. The dual costs are introduced in Section~\ref{sec:cg}.

\subsection{Dominance}\label{sec:dominance}

In this section, we introduce a dominance criterion used to mitigate the proliferation of labels. For two labels $L^1$ and $L^2$ that are attached to the same node, $L^1$ is said to dominate $L^2$ if: (i) the reduced cost of $L^1$ is better than the reduced cost of $L^2$, and (ii) any feasible extension of $L^2$ is either a. feasible for $L^1$, or b. dominated by an extension of $L^1$. 
%In other words, if each resource is less consumed in $L^1$ than in $L^2$. 
With the same notations as in the previous section, we can say that $L^1$ dominates $L^2$ if:

\begin{enumerate}
    \item $v^1 = v^2$;
    \item $c^1 \leq c^2 - \sum_{c \in \Theta_{2\backslash 1}} \sigma_c$, where $\Theta_{2 \backslash 1} = \{c \in \mathcal{C}^*\, |\, T^2_c = R,\, T^1_c \in \{V,U\}\}$ is the set of SL cuts that are reachable by an extension of $L^2$, but unreachable by an extension of $L^1$;
    \item For all $l \in \mathcal{L}$, if $R^2_l = R$ then $R^1_l = R$;
    \item $q^1 = q^2$;
    \item $F^1 = F^2$.
\end{enumerate}

Note that the number of edges $q$ included in the path, as well as the vector of mandatory stops $F$, must match to establish dominance between two labels. Since these two resources must be entirely consumed in valid solutions (i.e. the constraints are satisfied with equality), two labels with different consumption of these resources do not have the same possibilities of extension, and cannot be compared in terms of dominance. Condition 2 takes into account the potential dual contribution of the reachable SL-cuts and it is inspired by the work of \cite{jepsen_subset-row_2008}. Since the SL cuts are non-robust (see Section~\ref{sec:og_cuts}), they act as resources in the pricing procedure. \rev{Instead of naively comparing the dominance of $L^1$ and $L^2$, condition 2 proposes a stronger alternative where $L^2$ is discarded if every feasible extension would be dominated by a feasible extension of $L^1$. Indeed, even when accounting for the dual contribution of the cut-related resources that are consumed in $L^1$ and not in $L^2$, the cost of $L^2$ would still be dominated.}

\subsection{Pricing problem with other routing policies}\label{sec:pricing_other_policies}

The above subsections present the algorithm developed for the SLAPRP with the optimal routing policy, i.e. the case where the storage and routing problems are fully integrated. In the following, we will describe how to adapt the algorithm to other routing policies on a single block warehouse. In particular, we consider return, S-shape, midpoint, and largest gap. The only parts that are impacted are the underlying graph $\mathcal{G}$ and the resource extension function. Note that for some policies, additional parameters are required in the label definition to be able to compute the extension of resources.

\begin{figure}[h!]
    \centering
    \includegraphics{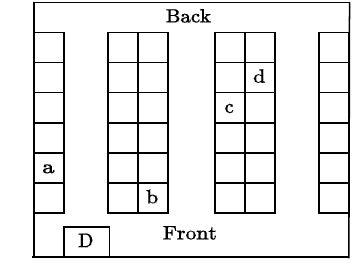}
    \caption{Illustrative example for heuristic routing policies.}
    \label{fig:pricing_example}
\end{figure}

\paragraph{Return routing.}
In the return routing policy, the distance between two locations is constant and does not depend on the locations previously visited by a partial path. The only required adaptation is to change the graph $\mathcal{G}$ to account for the modified distances and remove the arcs that are not feasible anymore. For example, in Figure~\ref{fig:pricing_example}, it is possible to go from locations a to b, but not the opposite. Note that the aisles are always visited in the same order, and the locations in an aisle are visited from the closest to the drop-off point to the farthest. In this case, it appears that the graph $\mathcal{G}$ is acyclic, which significantly reduces the computation burden of the algorithm, as the elementary aspect of the ESPPRC is guaranteed by the graph structure. 

\paragraph{S-shape routing.}
In the S-shape routing policy (as well as in those that follow), the distance between two locations is not constant. As shown in Figure~\ref{fig:pricing_example}, if we want to extend a label from location c to location d, the path followed by the picker depends on whether aisle 2 was entered by the front or the back cross aisle. Thus, an additional parameter is associated with each label, encoding if the current aisle was entered by the front or back cross aisle. When a label is extended to a new aisle, the value of the parameter is switched. The value of this parameter affects both: (i) The reachability of a location, for example, an extension from location b to c is only feasible if aisle 2 has been entered by the front; (ii) The distance between two locations, which depends on the orientation of the picker. Note that similarly to the return policy, the size of $\mathcal{G}$ is also reduced as some arcs become infeasible.

\paragraph{Midpoint routing.}

With the midpoint routing policy, the first and last aisles are crossed entirely. In the other aisles, the middle point is never crossed: the locations above it are visited by a path coming from the back cross aisle, and the locations below it are visited by a path coming from the front. If a single aisle is visited, a return policy is applied, in order not to create absurd routes. Therefore, the route development differs for the first and the last aisles. To account for this difference, two additional parameters are added to the label definition to register which aisle is the first (resp. last) one visited in the current path. The values of these two parameters are updated as follows:
\begin{itemize}
    \item \textit{First aisle:} The parameter is set during the first extension and is not modified afterward. All the locations in previous aisles are set unreachable in vector $R$.
    \item \textit{Last aisle:} The parameter is set during the first extension in a location that is below the midpoint, in an aisle that is not the first one. For example, in Figure~\ref{fig:pricing_example} we suppose that a label is attached to location a. If the label is extended to location c, above the midpoint, nothing happens. If it is extended to location b, below the midpoint, the second aisle is registered as the last one. Once the last aisle is set, some locations become unreachable, namely the ones above the midpoint in previous aisles, and all locations in the following aisles. For example, if a path visited locations a and b in Figure~\ref{fig:pricing_example}, location d would not be reachable anymore. If only one aisle is visited in the route, this parameter remains undefined.
\end{itemize}

\paragraph{Largest gap routing.}

The modeling of the largest gap policy is more intricate, due to the need to consider only feasible patterns in each aisle. The details are presented in Appendix~\ref{appendix:largest_pricing}.

\section{Computational experiments}\label{sec:experiments}

In this section, we report a summary of the experiments performed on the developed algorithm for the SLAPRP and its variants. The algorithm is coded in Julia 1.7.2, and CPLEX 12.10 is used to solve linear programs. We set CPLEX to use the dual simplex method, all other parameters use the default setting. The experiments are performed in {\em single-thread computation} on an Intel Xeon E5-2660 v3 with 10 cores at 2.6 GHz, with a 16 GB memory limit. In all the experiments, the integrality gap is computed as $100(\overline{z} - \underline{z})/\overline{z}$, where $\underline{z}$ is the value of the lower bound, and $\overline{z}$ the upper bound. All computing times are expressed in seconds. The detailed computational results, as well as the used instances, are available at \url{https://zenodo.org/record/7866860}. The source code for the implementation is available at \url{https://github.com/prunett/SLAPRP}\footnote{The repository will be uploaded upon publication}. In this section, we first describe the instances that are used to evaluate the performance of the procedures (see Section~\ref{sec:results:instances}) and then show the impact of the SL cuts (see Section~\ref{sec:results:formulation}) and the impact of the proposed branching rule (see Section~\ref{sec:result:branching}). We then compare our results with state-of-the-art results from the literature (see Sections~\ref{sec:results:silva}-\ref{sec:results:guo}).

\subsection{Benchmark instances}\label{sec:results:instances}

Two sets of instances from the literature are used in the computational experiments and are briefly described in the  following.

\paragraph{Instances of \cite{silva_integrating_2020}.} 
These instances use a classical single block layout, with the number of aisles in the range $\overline{a} \in \{1,3,5\}$, \rev{each of them comprising $\overline{b} \in \{5,10\}$ storage locations}. The number of SKUs is then comprised between 10 and 100. The demand is characterized by a number of orders $|\mathcal{O}| \in \{1,5,10\}$, each of them containing the same amount of products $|\mathcal{S}(o)| \in \{3,5\}$, sampled with a uniform distribution. A total of 108 instances are then solved with the classical routing policies, namely optimal, return, S-shape, midpoint, and largest gap. Section~\ref{sec:results:formulation} uses the full set of instances, with optimal routing. Since the instances with a single order can be solved by hand, and are all solved to optimality at the root node by our formulation, they are excluded from the benchmark set in Sections~\ref{sec:result:branching} and~\ref{sec:results:silva}.

\paragraph{Instances of \cite{guo_storage_2021}.}
These instances are largely inspired by a real-life warehouse. The studied problem is a variant of the SLAPRP that focuses on the replenishment problem. The set of fixed positions $\mathcal{F}$ is nonempty, more precisely there is $\alpha \in \{20\%,30\%,40\%\}$ of the SKUs whose locations are free, the other ones are already in stock in the picking area. \rev{The demand is modeled by a number of orders $|\mathcal{O}| \in \{50,100,200\}$, each of them comprising a random amount of SKUs in the range $|\mathcal{S}(o)| \in \{1,\dots,10\}$}. The layout is also inspired by the studied company, with a double-block layout with the drop-off point located at the beginning of the middle cross-aisle. \rev{A total of 80 SKUs are stored in the layout, and a} return routing policy is used.

\subsection{Impact of the SL cuts}\label{sec:results:formulation}

Table~\ref{tab:results:dual} reports on the quality of the different formulations for the SLAPRP considering the optimal routing. It also reports the impact of the SL inequalities on the quality of the lower bound. Results are obtained on the instances proposed in~\cite{silva_integrating_2020}. 
Column (LP) reports the integrality gap of the linear relaxation for the compact formulation \eqref{eq:c1:01}-\eqref{eq:c1:10}. \rev{(LP - MCF) presents the LP relaxation when using the multi-commodity flow formulation of Appendix~\ref{appendix:compact_mcf}.} (DW) reports the integrality gap of the extended formulation \eqref{eq:dw2:01}-\eqref{eq:dw2:06}. (DW + SL1) reports results obtained with the extended formulation strengthened with all and only SL-1 inequalities, while (DW + SL) reports results obtained with the extended formulation strengthened SL cuts of any order. Note that to obtain the results of Table~\ref{tab:results:dual} we decided to include an SL inequality only if it is violated by at least 0.01. 
%This is different from the Branch-Price-and-Cut algorithm where we are more conservative and only add an inequality violated by a threshold of 0.1. This is done to better highlight the potential of the SL inequalities. 
For each formulation, the column \textit{gap} reports the integrality gap. The column \textit{closed} shows how much the integrality gap is closed compared to the linear relaxation. The entries are computed as $100(\underline{z} - \underline{z}^{LP})/(\overline{z} - \underline{z}^{LP})$ where $\underline{z}$ is the current dual bound, $\underline{z}^{LP}$ the linear relaxation, and $\overline{z}$ the optimal (or best known) solution \rev{for the instance}.

\begin{table}[htp]
\centering
\caption{Impact of the formulation on the dual bound}
\label{tab:results:dual}
\begin{tabular}{llrrrrrrrrrrrrrrr}
\hline
\multicolumn{2}{l}{Instance} &   & \multicolumn{1}{l}{\hspace{.5cm}} & \multicolumn{1}{l}{LP} & \multicolumn{1}{l}{\hspace{.5cm}} & \multicolumn{2}{l}{\rev{LP - MCF}}                       & \multicolumn{1}{l}{\hspace{.5cm}} & \multicolumn{2}{l}{DW}                             & \multicolumn{1}{l}{\hspace{.5cm}} & \multicolumn{2}{l}{DW + SL1}                       & \multicolumn{1}{l}{\hspace{.5cm}} & \multicolumn{2}{l}{DW + SL}                        \\ \cline{7-8} \cline{10-11} \cline{13-14} \cline{16-17} 
Layout        & Orders       & $n$ & \multicolumn{1}{l}{}                               & \multicolumn{1}{l}{Gap}  & \multicolumn{1}{l}{}                               & \multicolumn{1}{l}{Gap} & \multicolumn{1}{l}{closed} & \multicolumn{1}{l}{}                               & \multicolumn{1}{l}{Gap} & \multicolumn{1}{l}{closed} & \multicolumn{1}{l}{}                               & \multicolumn{1}{l}{Gap} & \multicolumn{1}{l}{closed} & \multicolumn{1}{l}{}                               & \multicolumn{1}{l}{Gap} & \multicolumn{1}{l}{closed} \\ \hline

10 (1 x 5)& 3 (1 x 3) & 3 & & 50.0 & & \rev{33.2} & \rev{33.5} & & 0.0 & 100.0 & & 0.0 & 100.0 & & 0.0 & 100.0 \\ 
& 5 (1 x 5) & 3 & & 66.7 & & \rev{40.0} & \rev{40.0} & & 0.0 & 100.0 & & 0.0 & 100.0 & & 0.0 & 100.0 \\ 
& 15 (5 x 3) & 3 & & 67.0 & & \rev{15.3} & \rev{77.3} & & 10.6 & 84.8 & & 8.6 & 87.7 & & 5.2 & 92.5 \\ 
& 25 (5 x 5) & 3 & & 75.0 & & \rev{28.7} & \rev{61.8} & & 11.6 & 84.6 & & 9.2 & 87.8 & & 6.7 & 91.1 \\ 
& 30 (10 x 3) & 3 & & 69.5 & & \rev{14.1} & \rev{79.7} & & 17.2 & 75.2 & & 12.5 & 82.0 & & 5.6 & 91.9 \\ 
& 50 (10 x 5) & 3 & & 76.0 & & \rev{28.0} & \rev{63.2} & & 11.6 & 84.7 & & 10.1 & 86.7 & & 6.7 & 91.2 \\ 
\hline
20 (1 x 10)& 3 (1 x 3) & 3 & & 50.0 & & \rev{33.2} & \rev{33.5} & & 0.0 & 100.0 & & 0.0 & 100.0 & & 0.0 & 100.0 \\ 
& 5 (1 x 5) & 3 & & 66.7 & & \rev{40.0} & \rev{40.0} & & 0.0 & 100.0 & & 0.0 & 100.0 & & 0.0 & 100.0 \\ 
& 15 (5 x 3) & 3 & & 67.2 & & \rev{14.0} & \rev{79.1} & & 7.9 & 88.1 & & 6.7 & 90.0 & & 4.9 & 92.7 \\ 
& 25 (5 x 5) & 3 & & 81.0 & & \rev{20.6} & \rev{74.6} & & 14.1 & 82.6 & & 13.4 & 83.5 & & 7.1 & 91.3 \\ 
& 30 (10 x 3) & 3 & & 70.8 & & \rev{8.5} & \rev{88.0} & & 16.1 & 77.2 & & 13.1 & 81.4 & & 4.1 & 94.2 \\ 
& 50 (10 x 5) & 3 & & 81.5 & & \rev{20.2} & \rev{75.2} & & 20.2 & 75.3 & & 18.4 & 77.4 & & 9.2 & 88.8 \\ 
\hline
30 (3 x 5)& 3 (1 x 3) & 3 & & 50.0 & & \rev{33.2} & \rev{33.5} & & 0.0 & 100.0 & & 0.0 & 100.0 & & 0.0 & 100.0 \\ 
& 5 (1 x 5) & 3 & & 66.7 & & \rev{46.7} & \rev{30.0} & & 0.0 & 100.0 & & 0.0 & 100.0 & & 0.0 & 100.0 \\ 
& 15 (5 x 3) & 3 & & 59.2 & & \rev{20.9} & \rev{64.4} & & 6.9 & 88.6 & & 6.1 & 89.9 & & 4.2 & 93.0 \\ 
& 25 (5 x 5) & 3 & & 75.0 & & \rev{31.3} & \rev{58.2} & & 11.2 & 85.0 & & 11.1 & 85.1 & & 6.6 & 91.1 \\ 
& 30 (10 x 3) & 3 & & 63.2 & & \rev{21.5} & \rev{65.9} & & 17.7 & 71.9 & & 16.1 & 74.4 & & 12.1 & 80.9 \\ 
& 50 (10 x 5) & 3 & & 76.8 & & \rev{35.6} & \rev{53.7} & & 25.8 & 66.5 & & 25.3 & 67.1 & & 20.2 & 73.7 \\ 
\hline
60 (3 x 10)& 3 (1 x 3) & 3 & & 50.0 & & \rev{33.2} & \rev{33.5} & & 0.0 & 100.0 & & 0.0 & 100.0 & & 0.0 & 100.0 \\ 
& 5 (1 x 5) & 3 & & 66.7 & & \rev{46.7} & \rev{30.0} & & 0.0 & 100.0 & & 0.0 & 100.0 & & 0.0 & 100.0 \\ 
& 15 (5 x 3) & 3 & & 52.1 & & \rev{18.3} & \rev{65.0} & & 3.1 & 94.2 & & 3.1 & 94.2 & & 3.1 & 94.2 \\ 
& 25 (5 x 5) & 3 & & 69.3 & & \rev{27.2} & \rev{60.7} & & 5.1 & 92.7 & & 5.0 & 92.9 & & 3.5 & 95.0 \\ 
& 30 (10 x 3) & 3 & & 57.1 & & \rev{13.6} & \rev{76.1} & & 7.3 & 87.3 & & 6.2 & 89.2 & & 3.5 & 93.8 \\ 
& 50 (10 x 5) & 3 & & 76.2 & & \rev{24.5} & \rev{67.9} & & 17.3 & 77.3 & & 17.2 & 77.4 & & 11.9 & 84.4 \\ 
\hline
50 (5 x 5)& 3 (1 x 3) & 3 & & 50.0 & & \rev{33.2} & \rev{33.5} & & 0.0 & 100.0 & & 0.0 & 100.0 & & 0.0 & 100.0 \\ 
& 5 (1 x 5) & 3 & & 66.7 & & \rev{46.7} & \rev{30.0} & & 0.0 & 100.0 & & 0.0 & 100.0 & & 0.0 & 100.0 \\ 
& 15 (5 x 3) & 3 & & 53.2 & & \rev{21.2} & \rev{60.0} & & 4.4 & 91.9 & & 4.4 & 91.9 & & 4.2 & 92.3 \\ 
& 25 (5 x 5) & 3 & & 73.9 & & \rev{30.4} & \rev{58.9} & & 8.8 & 88.1 & & 8.6 & 88.4 & & 5.8 & 92.2 \\ 
& 30 (10 x 3) & 3 & & 57.6 & & \rev{14.8} & \rev{74.3} & & 8.9 & 84.6 & & 7.4 & 87.1 & & 4.0 & 93.0 \\ 
& 50 (10 x 5) & 3 & & 75.7 & & \rev{33.0} & \rev{56.5} & & 21.6 & 71.6 & & 21.5 & 71.7 & & 16.8 & 77.9 \\ 
\hline
100 (5 x 10)& 3 (1 x 3) & 3 & & 50.0 & & \rev{33.2} & \rev{33.5} & & 0.0 & 100.0 & & 0.0 & 100.0 & & 0.0 & 100.0 \\ 
& 5 (1 x 5) & 3 & & 66.7 & & \rev{46.7} & \rev{30.0} & & 0.0 & 100.0 & & 0.0 & 100.0 & & 0.0 & 100.0 \\ 
& 15 (5 x 3) & 3 & & 52.1 & & \rev{18.3} & \rev{65.0} & & 3.1 & 94.2 & & 3.1 & 94.2 & & 3.1 & 94.2 \\ 
& 25 (5 x 5) & 3 & & 71.2 & & \rev{30.1} & \rev{57.8} & & 8.5 & 88.2 & & 8.5 & 88.2 & & 5.8 & 91.9 \\ 
& 30 (10 x 3) & 3 & & 55.4 & & \rev{15.5} & \rev{72.0} & & 5.4 & 90.2 & & 5.2 & 90.6 & & 4.7 & 91.5 \\ 
& 50 (10 x 5) & 3 & & 71.7 & & \rev{23.1} & \rev{67.8} & & 9.2 & 87.2 & & 9.1 & 87.3 & & 6.2 & 91.4 \\ 

 \hline
\multicolumn{2}{l}{Total}    & 108  &  & 64.7 & & \rev{27.6} & \rev{56.2}  & & 7.6 & 89.2 & & 6.9 & 90.1  & & 4.6  & 93.4  \\ 
\multicolumn{3}{l}{\rev{Number of variables}} & \multicolumn{3}{r}{\rev{$|\mathcal{O}||\mathcal{L}|^2$}} & \multicolumn{3}{c}{\rev{$|\mathcal{O}||\mathcal{L}|^3$}} & \multicolumn{3}{c}{\rev{$|\mathcal{O}|e^{|\mathcal{L}|}$}} & \multicolumn{3}{c}{\rev{$|\mathcal{O}|e^{|\mathcal{L}|}$}} & \multicolumn{2}{c}{\rev{$|\mathcal{O}|e^{|\mathcal{L}|}$}} \\ \hline
\end{tabular}
\end{table}

The results show that the compact formulation presents a poor relaxation, with an average gap of 64.7\% over all the instances \rev{obtained by the compact formulation (2)--(12) when subtour constraints are expressed in the MTZ form, and 27.6\% when using the formulation reported in Appendix A where  subtour constraints are expressed in the MCF form.}. The DW reformulation is effective in tightening the formulation, with 89.2\% of the gap closed. Note that all the instances with a single order are solved to optimality by the reformulation. Concerning the SL inequalities, they prove to be effective by closing an additional 4.3\% gap when added to the DW formulation, which represents a closing of 39.5\% of the remaining gap. The SL-1 inequalities are also beneficial to strengthen the DW formulation, by closing 9.3\% of the integrality gap. These experiments show the strength of the proposed formulation, however, the integrality gap remains challenging to close for large instances.

\subsection{Impact of the branching schemes}\label{sec:result:branching}

This section reports on the impact of the different branching schemes on the size of the Branch-and-Bound tree. These experiments are performed on the instances of~\cite{silva_integrating_2020}, where we removed the instances with a single order. We consider the optimal routing policy and allow for 2 hours of computational time. Table~\ref{tab:results_tree} shows the results with 4 different settings: {\em Location} and {\em Combined} consider the respective branching schemes, while {\em Location + sym} and {\em Combined + sym} also take advantage of the symmetry breaking procedure presented in Section~\ref{sec:symmetry}. For each configuration, column \textit{opt} reports the number of instances solved to optimality. Columns \textit{gap}, \textit{nodes} and \textit{time} respectively report the average gap, the average number of nodes processed, and the average computing time for the instances not solved to optimality by all the configurations. Note that this is a fair comparison since the same instances are not solved to optimality on all the settings. Since the combined branching is equivalent to the location branching for instances with a single aisle (i.e., the first two layouts), we only report the results on one case on such instances in Table~\ref{tab:results_tree}.

%\begin{landscape}
\addtolength{\tabcolsep}{-1pt}
\begin{table}[h]
\caption{Impact of the branching scheme}
\label{tab:results_tree}
\small
%\resizebox{\columnwidth}{!}{
\makebox[\textwidth]{
\begin{tabular}{llrrrrrrrrrrrrrrrrrrrr}
\hline
\multicolumn{2}{l}{Instance} &  & \multicolumn{4}{c}{Location}                                                                                 & \multicolumn{1}{c}{\hspace{.5cm}} & \multicolumn{4}{c}{Location + sym}                                                                           & \multicolumn{1}{c}{\hspace{.5cm}} & \multicolumn{4}{c}{Combined}                                                                                   & \multicolumn{1}{c}{\hspace{.5cm}} & \multicolumn{4}{c}{Combined + sym}                                                                              \\ \cline{4-7} \cline{9-12} \cline{14-17} \cline{19-22} 
Layout        & Orders       &   $n$    & \multicolumn{1}{r}{opt} & \multicolumn{1}{r}{gap} & \multicolumn{1}{r}{nodes} & \multicolumn{1}{r}{time} & \multicolumn{1}{r}{}      & \multicolumn{1}{r}{opt} & \multicolumn{1}{r}{gap} & \multicolumn{1}{r}{nodes} & \multicolumn{1}{r}{time} & \multicolumn{1}{r}{}      & \multicolumn{1}{r}{opt} & \multicolumn{1}{r}{gap} & \multicolumn{1}{r}{nodes} & \multicolumn{1}{r}{time} & \multicolumn{1}{r}{}      & \multicolumn{1}{r}{opt} & \multicolumn{1}{r}{gap} & \multicolumn{1}{r}{nodes} & \multicolumn{1}{r}{time} \\ \hline

10 (1x5)& 15 (5x3) & 3 &3 & - & 39 & 1.9 & &3 & - & 41 & 1.9 & & -&- & -& -& &- &- &- &- \\ 
& 25 (5x5) &3 &3 & - & 209 & 18.1 & &3 & - & 176 & 18.9 & &- &- &- &- & &- &- & -& -\\ 
& 30 (10x3) & 3&3 & - & 28 & 3.4 & &3 & - & 30 & 3.4 & &- &- &- &- & &- &- &- &- \\ 
& 50 (10x5) &3 &3 & - & 455 & 38.1 & &3 & - & 450 & 37.1 & &- &- &- &- & &- &- &- &- \\ 
\hline
20 (1x10)& 15 (5x3) &3 &3 & - & 1 & 2.6 & &3 & - & 1 & 2.5 & &- &- &- &- & &- &- &- &- \\ 
& 25 (5x5) &3 &3 & - & 39449 & 1847.7 & &3 & - & 9286 & 354.9 & &- &- &- &- & &- &- &- &- \\ 
& 30 (10x3) &3 &3 & - & 57 & 8.7 & &3 & - & 61 & 8.8 & &- &- &- &- & &- &- &- &- \\ 
& 50 (10x5) & 3&2 & 1.5 & 8284 & 1620.3 & &2 & 1.5 & 7862 & 1436.0 & &- &- &- &- & &- &- & -&- \\ 
\hline
Total  & & 24 & 23 & 1.5 & 5669 & 391.4  & & 23 & 1.5   & 1994  & 180.6  & &- &- &- &- & & -& -& -&- \\
\hline
30 (3x5)& 15 (5x3) &3 &3 & - & 47 & 6.5 & &3 & - & 22 & 5.9 & &3 & - & 8 & 5.3 & &3 & - & 19 & 6.0 \\ 
& 25 (5x5) &3 &3 & - & 83 & 82.5 & &3 & - & 71 & 77.8 & &3 & - & 20 & 69.2 & &3 & - & 20 & 69.6 \\ 
& 30 (10x3) &3 &3 & - & 1427 & 183.3 & &3 & - & 1434 & 162.4 & &3 & - & 405 & 70.9 & &3 & - & 392 & 67.0 \\ 
& 50 (10x5) &3 &0 & 8.8 & - & - & &0 & 8.1 & - & - & &1 & 2.7 & - & - & &1 & 2.7 & - & - \\ 
\hline
60 (3x10)& 15 (5x3) &3 &3 & - & 68 & 17.1 & &3 & - & 21 & 13.5 & &3 & - & 13 & 11.8 & &3 & - & 13 & 12.0 \\ 
& 25 (5x5) &3 &3 & - & 37 & 90.8 & &3 & - & 18 & 86.1 & &3 & - & 15 & 81.4 & &3 & - & 14 & 79.6 \\ 
& 30 (10x3) &3 &3 & - & 430 & 142.8 & &3 & - & 479 & 120.2 & &3 & - & 384 & 140.4 & &3 & - & 285 & 103.0 \\ 
& 50 (10x5) &3 &0 & 13.9 & - & - & &0 & 11.5 & - & - & &0 & 2.9 & - & - & &0 & 2.9 & - & - \\ 
\hline
50 (5x5)& 15 (5x3) &3 &3 & - & 99 & 19.3 & &3 & - & 20 & 11.8 & &3 & - & 13 & 10.9 & &3 & - & 7 & 9.1 \\ 
& 25 (5x5) &3 &3 & - & 38 & 119.1 & &3 & - & 28 & 122.8 & &3 & - & 15 & 100.5 & &3 & - & 18 & 97.1 \\ 
& 30 (10x3) &3 &2 & - & 30 & 40.4 & &3 & - & 32 & 34.4 & &3 & - & 34 & 41.6 & &3 & - & 34 & 41.1 \\ 
& 50 (10x5) &3 &0 & 13.3 & - & - & &0 & 11.7 & - & - & &0 & 6.4 & - & - & &0 & 6.5 & - & - \\ 
\hline
100 (5x10)& 15 (5x3) &3 &3 & - & 109 & 73.5 & &3 & - & 36 & 39.0 & &3 & - & 11 & 29.1 & &3 & - & 11 & 29.2 \\ 
& 25 (5x5) &3 &3 & - & 86 & 679.7 & &3 & - & 60 & 674.3 & &3 & - & 26 & 457.7 & &3 & - & 14 & 307.9 \\ 
& 30 (10x3) &3 &0 & 2.5 & - & - & &1 & 2.5 & - & - & &0 & 2.5 & - & - & &0 & 2.5 & - & - \\ 
& 50 (10x5) &3 &0 & 7.2 & - & - & &0 & 5.8 & - & - & &0 & 4.1 & - & - & &0 & 4.2 & - & - \\

\hline
Total  & & 48 & 32 & 9.7 & 229  & 135.2  & & 34 & 8.3 & 207  & 125.3  &   & 34  & 3.9 & 87 & 94.2 & & 34 & 3.9 & 76 & 75.7 \\ \hline
\end{tabular}}
\end{table}
%\end{landscape}

\addtolength{\tabcolsep}{1pt}

The obtained results show that the combined branching scheme increases the number of solved instances, and provides lower optimality gaps for the non-solved instances. For the instances with more than one aisle, the average tree size for closed instances is 229 for location branching, and 87 for combined branching, i.e., a reduction of 62\%. The gap for non-solved instances is also reduced from an average of 9.7 to 3.9 when using the combined strategy. The strengthening procedure appears to be effective in limiting the size of the search tree, with the average number of processed nodes reduced from 5669 to 1994 for the instances with a single aisle, i.e., a reduction of 65\%. For the instances with multiple aisles, the number of processed nodes is reduced by 9\% for location branching and 13\% for combined branching. Overall the combined scheme with the strengthening procedure is the most performing branching procedure.

\subsection{Comparison with Silva et al. (2020)}\label{sec:results:silva}

In this section, we test our algorithm on the benchmark instances of \cite{silva_integrating_2020} with more than one order. The authors propose compact formulations for the SLAPRP with a single block warehouse and its variants, where the routing aspect of the problem is tackled by heuristic policies (i.e., return, S-shape, midpoint, and largest gap). Table~\ref{tab:silva_benchmark} presents the results of three solution methods. \textit{BCP} designates the Branch-Cut-and-Price algorithm. \textit{CPLEX (Silva et al.)} presents the results obtained by solving the MILP formulations introduced by \cite{silva_integrating_2020} using CPLEX. \textit{CPLEX (Prunet et al.)} shows the results obtained by solving our MILP formulations using CPLEX. Note that our formulations have several differences with those of \cite{silva_integrating_2020}: i. For the optimal routing, we use formulation (MCF) instead of the cubic formulation of \cite{silva_integrating_2020}. ii. For the midpoint policy, the set of rules defining the policy in \cite{silva_integrating_2020} is more intricate than what is commonly done in the literature \citep{de_koster_design_2007}. Therefore, we propose a simpler modeling, consistent with the order-picking literature. We stress that, for this policy, the studied problem is not the same as in \cite{silva_integrating_2020}, and the results cannot be compared. iii. For the other policies, \cite{silva_integrating_2020} propose non-linear formulations, that use quadratic terms and if-then-else constraints. In Appendices~\ref{appendix:compact_return}--\ref{appendix:compact_largest}, we propose linearized alternative formulations only using indicator constraints for conditional statements. For each solution method, we report (\textit{opt}) the number of instances solved to optimality, the average gap (\textit{gap}), and the average computing time (\textit{time}). Note that the time limit for computation is fixed to 2 hours. For the BCP, we also report the average number of processed nodes (\textit{nodes}), and generated SL cuts (\textit{cuts}). The column $\Delta_{opt}$ computes the average gap between the best-known solution for the routing policy at hand, and the optimal routing, computed as $100(\overline{z}^{policy} - \overline{z}^{opt})/\overline{z}^{opt}$. To ensure a fair comparison, we apply to the CPLEX runs the post-processing routine introduced in Section~\ref{sec:branching} (Implementation details) to strengthen the lower bounds.

\begin{table}[htp]
\caption{Results on the benchmark set of \cite{silva_integrating_2020}}
\label{tab:silva_benchmark}
\small
\makebox[\textwidth]{
\begin{tabular}{llrrrrrrrrrrrrrrrrr}
\hline
\multicolumn{2}{l}{Instance} & \multicolumn{1}{l}{} & \multicolumn{1}{l}{\hspace{.5cm}} & \multicolumn{1}{l}{}      & \multicolumn{1}{l}{\hspace{.5cm}} & \multicolumn{3}{c}{CPLEX (Silva et al.)} & \multicolumn{1}{l}{\hspace{.5cm}} & \multicolumn{3}{c}{CPLEX (Prunet et al.)} & \multicolumn{1}{l}{\hspace{.5cm}} & \multicolumn{5}{c}{BCP}                \\ \cline{7-9} \cline{11-13} \cline{15-19} 
Routing policy  & Layout     & $n$                    &                            & \multicolumn{1}{l}{$\Delta_{opt}$} & \multicolumn{1}{l}{}       & opt     & gap      & time      &                            & opt     & gap      & time       &                            & opt & gap  & time   & nodes & cuts  \\ \hline

Optimal & 10 (1x5) & 12 & & - & & 0 & 92.0 & 6163.4 & & \rev{12} & \rev{0.0} & \rev{17.8} & & 12 & 0.0 & 15.1 & 174 & 73 \\ 
& 20 (1x10) & 12 & & - & & 0 & 100.0 & 7200.1 & & \rev{10} & \rev{0.9} & \rev{1941.0} & & 11 & 0.1 & 910.8 & 5645 & 375 \\ 
& 30 (3x5) & 12 & & - & & 0 & 100.0 & 7122.3 & & \rev{9} & \rev{11.5} & \rev{3618.6} & & 10 & 0.5 & 1730.5 & 1172 & 605 \\ 
& 60 (3x10) & 12 & & - & & 0 & 100.0 & 7823.1 & & \rev{3} & \rev{26.0} & \rev{5769.1} & & 9 & 0.7 & 1848.7 & 262 & 1397 \\ 
& 50 (5x5) & 12 & & - & & 0 & 100.0 & 7214.5 & & \rev{3} & \rev{15.9} & \rev{5544.3} & & 9 & 1.6 & 1992.8 & 966 & 938 \\ 
& 100 (5x10) & 12 & & - & & 0 & 100.0 & 7273.4 & & \rev{3} & \rev{31.7} & \rev{5955.9} & & 6 & 1.7 & 3684.4 & 1597 & 586 \\ 
Subtotal & & 72 & & - & & 0 & 98.3 & 7089.6 & & \rev{40} & \rev{14.3} & \rev{3807.8} & & 57 & 0.8 & 1697.0 & 1636 & 662 \\ \hline 
Return & 10 (1x5) & 12 & & 0.0 & & 12 & 0.0 & 0.1 & & 12 & 0.0 & 5.2 & & 12 & 0.0 & 3.5 & 174 & 73 \\ 
& 20 (1x10) & 12 & & 0.0 & & 12 & 0.0 & 6.4 & & 12 & 0.0 & 8.4 & & 11 & 0.1 & 884.1 & 5677 & 375 \\ 
& 30 (3x5) & 12 & & 0.9 & & 9 & 3.5 & 1826.1 & & 9 & 3.6 & 1825.9 & & 12 & 0.0 & 1423.2 & 1978 & 644 \\ 
& 60 (3x10) & 12 & & 0.0 & & 9 & 3.9 & 1863.3 & & 9 & 3.6 & 1906.6 & & 10 & 0.4 & 1479.4 & 568 & 1273 \\ 
& 50 (5x5) & 12 & & 0.0 & & 8 & 5.7 & 2474.7 & & 8 & 6.0 & 2496.8 & & 9 & 1.0 & 1864.1 & 1800 & 925 \\ 
& 100 (5x10) & 12 & & 0.0 & & 6 & 5.4 & 3684.2 & & 6 & 7.8 & 3865.5 & & 9 & 0.6 & 2663.3 & 5956 & 557 \\ 
Subtotal & & 72 & & 0.1 & & 56 & 3.1 & 1642.5 & & 56 & 3.5 & 1684.7 & & 63 & 0.4 & 1386.3 & 2692 & 641 \\ \hline 
S-Shape & 10 (1x5) & 12 & & 0.0 & & 12 & 0.0 & 0.1 & & 12 & 0.0 & 6.9 & & 12 & 0.0 & 4.2 & 174 & 73 \\ 
& 20 (1x10) & 12 & & 0.0 & & 12 & 0.0 & 6.0 & & 12 & 0.0 & 13.8 & & 11 & 0.1 & 889.1 & 5655 & 375 \\ 
& 30 (3x5) & 12 & & 13.3 & & 12 & 0.0 & 180.8 & & 12 & 0.0 & 137.7 & & 11 & 0.1 & 1696.5 & 19673 & 1261 \\ 
& 60 (3x10) & 12 & & 14.6 & & 9 & 0.7 & 1880.7 & & 9 & 1.0 & 1971.4 & & 7 & 1.3 & 3519.0 & 9752 & 10170 \\ 
& 50 (5x5) & 12 & & 9.1 & & 9 & 1.6 & 2041.4 & & 10 & 0.6 & 1647.6 & & 10 & 0.4 & 2637.9 & 9373 & 1666 \\ 
& 100 (5x10) & 12 & & 9.2 & & 9 & 1.4 & 1920.5 & & 9 & 2.0 & 2175.5 & & 6 & 1.4 & 4145.5 & 4734 & 1210 \\ 
 Subtotal & & 72 & & 5.1 & & 63 & 0.6 & 1004.9 & & 64 & 0.6 & 992.2 & & 57 & 0.6 & 2148.7 & 8227 & 2459 \\ \hline 
Midpoint* & 10 (1x5) & 12 & & 0.0 & & 12 & 0.0 & 0.7 & & 12 & 0.0 & 6.8 & & 12 & 0.0 & 4.3 & 174 & 73 \\ 
& 20 (1x10) & 12 & & 0.0 & & 12 & 0.0 & 56.7 & & 12 & 0.0 & 13.9 & & 11 & 0.1 & 885.1 & 5668 & 375 \\ 
& 30 (3x5) & 12 & & 13.1 & & 6 & 28.3 & 3933.9 & & 12 & 0.0 & 236.8 & & 11 & 0.2 & 1405.0 & 17163 & 1380 \\ 
& 60 (3x10) & 12 & & 15.1 & & 4 & 37.9 & 5176.3 & & 7 & 2.4 & 3295.9 & & 7 & 1.6 & 3537.0 & 8089 & 8956 \\ 
& 50 (5x5) & 12 & & 9.3 & & 4 & 41.6 & 5375.8 & & 8 & 1.4 & 2639.3 & & 8 & 1.0 & 2802.9 & 7890 & 2569 \\ 
& 100 (5x10) & 12 & & 9.9 & & 1 & 47.4 & 6848.7 & & 6 & 5.8 & 3748.2 & & 5 & 2.3 & 4355.3 & 3029 & 5262 \\ 
Subtotal & & 72 & & 5.3 & & 39 & 25.9 & 3565.4 & & 57 & 1.6 & 1656.8 & & 54 & 0.9 & 2164.9 & 7002 & 3102 \\ \hline 
Largest gap & 10 (1x5) & 12 & & 0.0 & & 12 & 0.0 & 2.1 & & 12 & 0.0 & 7.4 & & 12 & 0.0 & 4.3 & 174 & 73 \\ 
& 20 (1x10) & 12 & & 0.0 & & 10 & 0.9 & 1758.3 & & 12 & 0.0 & 45.3 & & 11 & 0.1 & 885.8 & 5682 & 375 \\ 
& 30 (3x5) & 12 & & 13.1 & & 7 & 3.2 & 3099.7 & & 12 & 0.0 & 962.8 & & 12 & 0.0 & 1379.2 & 17212 & 1357 \\ 
& 60 (3x10) & 12 & & 14.9 & & 5 & 10.1 & 4741.9 & & 7 & 3.0 & 3200.3 & & 7 & 1.5 & 3512.1 & 8146 & 10761 \\ 
& 50 (5x5) & 12 & & 9.1 & & 6 & 15.9 & 3733.8 & & 7 & 2.9 & 3113.3 & & 8 & 0.9 & 2790.1 & 8298 & 2195 \\ 
& 100 (5x10) & 12 & & 9.5 & & 4 & 33.6 & 5384.4 & & 5 & 5.8 & 4499.2 & & 5 & 1.9 & 4319.1 & 3804 & 2570 \\ 
Subtotal & & 72 & & 5.2 & & 44 & 10.6 & 3120.0 & & 55 & 1.9 & 1971.4 & & 55 & 0.7 & 2148.4 & 7219 & 2888 \\ \hline 
\hline Total & & 360 & & 3.9 & & 202 & 27.7 & 3284.5 & & 272 & 4.4 & 2022.6 & & 286 & 0.7 & 1909.1 & 5355 & 1951 \\ \hline

\end{tabular}}
\end{table}

The BCP algorithm shows promising results on this benchmark set. For the completely integrated problem, where the routing is solved to optimality, the BCP clearly outperforms CPLEX with 57 instances solved to optimality, compared to none from the formulation of \cite{silva_integrating_2020}, and 40 for our compact formulation. For the other routing policies, the results are more nuanced: The BCP performs better than CPLEX on the number of instances closed to optimality with the return routing, worse on S-shape routing, and similarly on the midpoint and largest gap routing. These results are unsurprising, as the S-shape policy is a very straightforward heuristic that simplifies drastically the structure of the problem when leveraged with an ad hoc formulation. Note, however, that for all policies the BCP provides better gaps on average than the compact formulations, even when closing fewer instances to optimality. \rev{There are actually only 14 instances out of 272 where CPLEX returns a better dual bound at the end of the run, mostly with midpoint and S-Shape policies. Experiments show that CPLEX seems to perform better at finding good quality primal bounds, but struggles to close the gap when the initial dual bound is far from the optimum.} Overall on the whole set, we found 115 new optimal solutions out of 162 previously unsolved instances, without counting the ones for the midpoint policy. Looking at the column $\Delta_{opt}$, it appears that solving the routing problem to optimality leads to savings of around 5\% compared to the S-shape, midpoint, and largest gap policies. For the return policies, the results are very similar. This is due to the small size of the instances that leads to the optimal routing behaving similarly to the return routing in most cases. %making this experiment not conclusive for instances of small size.

\subsection{Comparison with Guo et al. (2021)}\label{sec:results:guo}

Table~\ref{tab:results_guo} shows the results of different algorithms on the benchmark instances of \cite{guo_storage_2021}. The first three columns characterize the instance, with the proportion of free assignments $\alpha$, the number of orders, and $n$ the number of instances of each type. Then we present the results of three solution methods: the MILP formulation introduced by \cite{guo_storage_2021}, dedicated to their specific instance of the SLAPRP and solved using CPLEX, a dynamic programming algorithm (DP) developed by \cite{guo_storage_2021} and our BCP algorithm. For each algorithm, we report the number of instances solved to optimality (\textit{opt}), and the computing time (\textit{time}). The results show that CPLEX hardly scales with the size of the problem in this configuration, and runs out of memory on 40 of the 90 instances. The DP and BCP algorithms both close all the instances to optimality. In terms of running time, both algorithms perform equivalently on the small instances ($\alpha = 0.2$), but the BCP clearly outperforms the DP by several orders of magnitude when scaling to larger instances with $\alpha = 0.4$. We believe this is due to the increased combinatorics of the large instances. 
For small values of $\alpha$, most of the storage plan is already fixed, and the enumeration of the feasible solution space is still tractable (e.g., for $\alpha = 0.2$ the total number of solutions is $8! = 40 320$), despite a challenging objective function with up to 200 orders, which is not the case for large instances. Note that the BCP algorithm closes all the instances at the root node, avoiding the burden of enumeration in this case.

\begin{table}[h]
\centering
\caption{Results for the benchmark set of \cite{guo_storage_2021}}
\label{tab:results_guo}
\begin{tabular}{lrrrrrrrrrrr}
\hline
\multicolumn{2}{l}{Instance}                   & \multicolumn{1}{l}{}  & \multicolumn{1}{l}{\hspace{.5cm}} & \multicolumn{2}{l}{CPLEX}                             & \multicolumn{1}{l}{\hspace{.5cm}} & \multicolumn{2}{l}{DP}                               & \multicolumn{1}{l}{\hspace{.5cm}} & \multicolumn{2}{l}{BCP}                              \\ \cline{5-6} \cline{8-9} \cline{11-12} 
$\alpha$ & \multicolumn{1}{l}{Orders} & \multicolumn{1}{l}{$n$} & \multicolumn{1}{l}{}                               & \multicolumn{1}{l}{\#opt} & \multicolumn{1}{l}{time} & \multicolumn{1}{l}{}                               & \multicolumn{1}{l}{\#opt} & \multicolumn{1}{l}{time} & \multicolumn{1}{l}{}                               & \multicolumn{1}{l}{\#opt} & \multicolumn{1}{l}{time} \\ \hline
0.2 & 50 & 10 & & 10 & 1.5 & & 10 & 0.9 & & 10 & 1.9 \\
0.2 & 100 & 10 & & 10 & 9.3 & & 10 & 1.4 & & 10 & 3.4 \\
0.2 & 200 & 10 & & 10 & 30.8 & & 10 & 2.5 & & 10 & 6.6 \\
0.3 & 50 & 10 & & 10 & 552.8 & & 10 & 34.4 & & 10 & 3.2 \\
0.3 & 100 & 10 & & 10 & 2484.8 & & 10 & 35.7 & & 10 & 6.2 \\
0.3 & 200 & 10 & & 0 & memory & & 10 & 44.4 & & 10 & 11.8 \\
0.4 & 50 & 10 & & 0 & memory & & 10 & 2856.6 & & 10 & 5.3 \\
0.4 & 100 & 10 & & 0 & memory & & 10 & 2967.2 & & 10 & 9.6 \\
0.4 & 200 & 10 & & 0 & memory & & 10 & 3247.5 & & 10 & 18.3 \\ \hline
\multicolumn{2}{l}{Total}                      & \multicolumn{1}{l}{90}  & \multicolumn{1}{l}{}                               & \multicolumn{1}{r}{50}       & \multicolumn{1}{r}{615.8*}     & \multicolumn{1}{l}{}                               & \multicolumn{1}{r}{90}      & \multicolumn{1}{r}{1021.2}     & \multicolumn{1}{l}{}                               & \multicolumn{1}{r}{90}      & \multicolumn{1}{r}{7.4}     \\ \hline
\end{tabular}
\end{table}

\section{Conclusion}\label{sec:conclusion}

In this paper, we introduced a new exact solution approach to solve the SLAPRP and a large class of its variants, including other warehouse layouts, the most common heuristic routing policies (i.e., return, S-shape, midpoint, and largest gap), and the partial replenishment variant of the problem. The developed approach is based on a Dantzig-Wolfe decomposition, where operational considerations are convexified in the pricing subproblems. Adapting our algorithm to a new variant of the problem only requires minor adjustments of the underlying graph and the resource extension function used in the pricing routine. The formulation is further tightened by the introduction of a new family of non-robust valid inequalities. The problem structure is exploited by the introduction of a novel branching scheme, further strengthened by a symmetry-breaking routine. These two contributions lead to a significant reduction in the Branch-and-Bound tree size (more than two thirds on average). The developed Branch-Cut-and-Price algorithm is benchmarked on the instances from \cite{silva_integrating_2020}, solving to optimality 115 previously unclosed instances. For the instances from \cite{guo_storage_2021}, which were already closed, our BCP algorithm scales better with the number of orders, being on large instances several orders of magnitude faster than the previous state-of-the-art algorithm. 

In the future, we expect further research on integrated problems in warehousing logistics to answer the challenges of e-commerce, as it has been highlighted by other authors \citep{boysen_warehousing_2019,van_gils_designing_2018}. A natural extension of the present work is the development of a matheuristic able to tackle industrial-scale instances as well as extending our algorithm to other variants of the SLAPRP (e.g., several block warehouses), which is left for future research. In Section~\ref{sec:decomposition}, we briefly discuss alternative decomposition methods for the SLAPRP formulation (i.e., Benders decomposition and variable splitting) that could lead to further methodological improvements for the problem. Finally, the integration of the batching decisions, which has been disregarded in the present work, alongside the storage and routing optimization is a promising direction for future research, which can further improve the understanding of the interactions between decision problems in the picking area.

% 152 closed instances in total (with the ones order = 1), 114 without the ones. These number counts the midpoint instances !!!

% Acknowledgments here
\subsection*{Acknowledgments}
This work has been supported by the French National Research Agency through the AGIRE project under the grant ANR-19-CE10-0014\footnote{\url{https://anr.fr/Projet-ANR-19-CE10-0014}}. This support is gratefully acknowledged. The authors thank prof. Valeria Borodin for useful feedback on the problem definition and comments on an earlier draft of this paper.

% Bibliography
%\newpage

%\bibliographystyle{informs2014} % outcomment this and next line in Case 1
%\bibliography{references} % if more than one, comma separated
%\printbibliography
\bibliographystyle{apalike} 
\bibliography{references_1}

\setstretch{1}
\appendix
\section{Integrality of the assignment polytope}\label{appendix:assignment_polytope}

In this section, we prove that the polytope consisting of the assignment variables and constraints is integral. This result is not surprising since its structure is very close to the classical assignment polytope.

\begin{proposition}\label{prop:assignment_integral}
    Let us define the polytope $\mathcal{P} = \{\xi_{ls} \in \mathbb{R}^{|\mathcal{S}|\cdot|\mathcal{L}|} \,|\, \eqref{eq:c1:01},\, \eqref{eq:c1:02}, \, 0 \leq \xi_{ls} \leq 1 \, \forall l \in \mathcal{L}, s \in \mathcal{S}\}$. If $\mathcal{P} \neq \emptyset$, i.e. the set of fixed assignments $\mathcal{F}$ is feasible, then $\mathcal{P}$ is integral, i.e. its extreme points are all integer.
\end{proposition}

The proof of this proposition relies largely on the following proposition:

\begin{proposition}\label{prop:nemhauser_integer_1988}[\cite{nemhauser_integer_1988}, III.1.2.3]
    If A is TU (i.e. Totally Unimodular), if b, b', d and d' are integral, and if $P(b,b',d,d') = \{x \in \mathbb{R}^n \, |\, b' \leq Ax\leq b,\, d' \leq x \leq d\}$ is not empty, then $P(b,b',d,d')$ is an integral polyhedron.
\end{proposition}

\paragraph{Proof of Proposition \ref{prop:assignment_integral}} The description of $\mathcal{P}$ matches the description of $P(b,b',d,d')$ given in Proposition \ref{prop:nemhauser_integer_1988} since the coefficients in all the constraints are integer, and $\mathcal{P}$ is not empty. Now, let us consider the matrix of constraints \eqref{eq:c1:01} and \eqref{eq:c1:02}. Its coefficients are all either 0 or 1. Moreover, for a given $l \in \mathcal{L}$ and $s \in \mathcal{S}$, the variable $\xi_{ls}$ appears exactly in two constraints: one in the set of constraints~\eqref{eq:c1:01}, and one in the set~\eqref{eq:c1:02}. Therefore, one column of the constraint matrix possesses exactly two coefficients 1, one in each block of constraints. It follows that the constraint matrix of $\mathcal{P}$ satisfies the Hoffman and Gale sufficient conditions to be TU. Therefore we proved that $\mathcal{P}$ is integral.
\begin{flushright}
$\square$
\end{flushright}
\section{Proof of Theorem \ref{prop:sl}}\label{appendix:sl_proof}

In this section, we prove that the SL inequalities are valid for formulation (DW).

\textbf{Proof of Theorem~\ref{prop:sl}}

Let us select $o \in \mathcal{O}$, $s \in \mathcal{S}(o)$ and $\overline{\mathcal{L}} \subset \mathcal{L}$. Let $(\xi,\rho)$ be an integer solution of (DW). According to the assignment constraints~\eqref{eq:dw2:02}, there is a unique location $l \in \mathcal{L}$ such that $\xi_{ls} = 1$, and $\xi_{hs} = 0$ for all $h \in \mathcal{L} \backslash \{l\}$. The two following cases are faced:
\begin{itemize}
    \item $l \not\in \overline{\mathcal{L}}$. In this case the right-hand side of inequality~\eqref{eq:el_cut} is equal to 0, and since the left-hand side is a sum of binary variables, multiplied by indicator coefficients, the inequality is valid.
    \item $l \in \overline{\mathcal{L}}$. In this case, the right-hand side of inequality~\eqref{eq:el_cut} is equal to 1, and we need to prove that the left-hand side is larger than or equal to 1. We note that, for a route $r \in \mathcal{R}_o$, we have $\delta_r(\{l\}) \geq \frac{a^{l}_{or}}{K_l}$ since $0 \leq a_{or}^l \leq K_l$, and $a_{or}^l \geq 1$ implies that $r$ visits $l$, and by extension $\overline{\mathcal{L}}$, so $\delta_r(\{l\}) = 1$. Then, using constraints~\eqref{eq:dw2:04} we have:
    $$
        \sum_{r \in \mathcal{R}_o} \delta_r(\overline{\mathcal{L}}) \rho_{or} \geq \sum_{r \in \mathcal{R}_o} \delta_r(\{l\}) \rho_{or}
        \geq \sum_{r \in \mathcal{R}_o} \frac{a_{or}^{l}}{K_l}\rho_{or}
        \geq \frac{1}{K_l} \sum_{s \in \mathcal{S}(o)} \xi_{ls}
        \geq \frac{1}{K_l} \xi_{ls} 
        \geq \frac{1}{K_l}$$
    Since the left-hand side is integer, as a sum of integer variables multiplied by indicator coefficients, we can strengthen this inequality to:
    \begin{equation*}
        \sum_{r \in \mathcal{R}_o} \delta_r(\overline{\mathcal{L}}) \rho_{or} \geq 1
    \end{equation*}
    Therefore the inequalities~\eqref{eq:el_cut} are valid.
\end{itemize}
\begin{flushright}
$\square$
\end{flushright}
\section{Proof of Theorem \ref{prop:branching}}\label{appendix:branching_proof}

We start the section by introducing some definitions. For a solution $(\xi,\rho)$ of $RMP(\overline{\mathcal{R}},\overline{\mathcal{C}}))$ and an order $o \in \mathcal{O}$, we call the \textit{support of o} the set of locations $Supp_o(\xi) = \{l \in \mathcal{L} \, ; \, \sum_{s\in \mathcal{S}(o)} \xi_{ls} > 0\}$. In other words, the support of $o$ with respect to the solution $(\xi,\rho)$ is the subset of locations where the SKUs of $o$ are stored, at least partially, in the solution.

\begin{lemma}\label{lemma1}
Let $(\xi,\rho)$ be a solution of $RMP(\overline{\mathcal{R}},\overline{\mathcal{C}})$, $o \in \mathcal{O}$ an order, and $r^* \in \mathcal{R}_o$ a valid route for $o$. If $\rho_{or^*} > 0$ then $\mathcal{L}(r^*) \subset Supp_o(\xi)$.
\end{lemma}

In other words, this lemma means that the chosen routes to collect an order $o\in\mathcal{O}$ stop in the locations belonging to the support of the corresponding order.

\textbf{Proof of Lemma~\ref{lemma1}}

We will prove this lemma by contraposition. We suppose that $\mathcal{L}(r^*) \not\subset Supp_o(\xi)$. We then prove that $\rho_{or^*} = 0$, which constitutes a sufficient proof.

Since $\mathcal{L}(r^*) \not\subset Supp_o(\xi)$, there exists a location $l^*$ that is visited by $r^*$ (i.e. $l^* \in \mathcal{L}(r^*)$) and not in the support of $o$ w.r.t. $\xi$ (i.e. $\sum_{s \in \mathcal{S}(o)} \xi_{l^*s} = 0$). We start by summing over $\mathcal{L}\backslash \{l^*\}$ the linking constraints~\eqref{eq:rmp:02} of the master problem:

\begin{equation}\label{eq:branching:proof01}
\sum_{l \in \mathcal{L}\backslash \{l^*\}} \sum_{s \in \mathcal{S}(o)} \xi_{ls} \leq \sum_{l \in \mathcal{L}\backslash \{l^*\}} \sum_{r \in \mathcal{R}_o} a_{or}^l \rho_{or}
\end{equation}

Since $l^*$ is not part of the support of $o$, we can transform the left hand side of the inequality~\eqref{eq:branching:proof01}, using the assignment constraint~\eqref{eq:rmp:05}:

$$\sum_{l \in \mathcal{L}\backslash \{l^*\}} \sum_{s \in \mathcal{S}(o)} \xi_{ls} = \sum_{l \in \mathcal{L}} \sum_{s \in \mathcal{S}(o)} \xi_{ls} = \sum_{s \in \mathcal{S}(o)} \sum_{l \in \mathcal{L}} \xi_{ls} = \sum_{s \in \mathcal{S}(o)} 1 = |\mathcal{S}(o)| $$

Concerning the right-hand side of inequality~\eqref{eq:branching:proof01}, we transform it as follows:

$$ \sum_{l \in \mathcal{L}\backslash \{l^*\}} \sum_{r \in \mathcal{R}_o} a_{or}^l \rho_{or} = \sum_{r \in \mathcal{R}_o} \sum_{l \in \mathcal{L}\backslash \{l^*\}}  a_{or}^l \rho_{or} = \left(\sum_{r \in \mathcal{R}_o\backslash \{r^*\}} \rho_{or} \sum_{l \in \mathcal{L}\backslash \{l^*\}}  a_{or}^l\right)  + \left(\rho_{or^*}\sum_{l \in \mathcal{L}\backslash \{l^*\}}  a_{or^*}^l\right)$$

Then we note that, for $r \in \mathcal{R}_o \backslash \{r^*\}$, $(\sum_{l \in \mathcal{L}\backslash \{l^*\}}  a_{or}^l) \leq (\sum_{l \in \mathcal{L}}  a_{or}^l) = |\mathcal{S}(o)|$. Indeed, the $a$ coefficients count, for a given route, the number of stops it makes at a given location. Therefore its sum over all locations corresponds to the total number of stops of a route, which is equal to the number of SKUs that are part of the corresponding order. %This is expressed as constraints~\eqref{eq:c1:06} in the compact formulation, which is then convexified and enforced in the pricing problems.
Concerning the second term, we remark that $\sum_{l \in \mathcal{L}\backslash \{l^*\}}  a_{or^*}^l  = |\mathcal{S}(o)|\, - a_{or^*}^{l^*} \leq |\mathcal{S}(o)| - 1$, since the starting hypothesis of this proof is that $r^*$ stops in $l^*$ (i.e. $a_{or^*}^{l^*} = 1$). We can then rewrite the inequality~\eqref{eq:branching:proof01} as:

\begin{equation}\label{eq:branching:proof02}
    |\mathcal{S}(o)|  \leq  |\mathcal{S}(o)| \sum_{r \in \mathcal{R}_o \backslash \{r^*\}} \rho_{or} + (|\mathcal{S}(o)| -1) \rho_{or^*}
\end{equation}

According to constraints~\eqref{eq:rmp:01}, we have $\sum_{r \in \mathcal{R}_o \backslash \{r^*\}} \rho_{or} = 1 - \rho_{or^*}$. By replacing this result in the inequality~\eqref{eq:branching:proof02} we obtain:

$$ |\mathcal{S}(o)| \quad \leq \quad |\mathcal{S}(o)| (1 - \rho_{or^*}) + (|\mathcal{S}(o)| -1) \rho_{or^*} \quad = \quad |\mathcal{S}(o)| - \rho_{or^*}$$

By rearranging the terms of the inequality we obtain $\rho_{or^*} \leq 0$, which terminates this proof.
\begin{flushright}
$\square$
\end{flushright}

\textbf{Proof of Theorem~\ref{prop:branching}}\\
Let $ X = (\xi,\rho)$ be a solution of $(RMP(\overline{\mathcal{R}},\overline{\mathcal{C}}))$ with all $\xi$ variables integer. We build a solution $X' = (\xi',\rho')$, with:
\begin{itemize}
    \item For all $l \in \mathcal{L}$ and $s \in \mathcal{S}$, $\xi_{ls}' = \xi_{ls}$.
    \item For an order $o \in \mathcal{O}$, we build the route $r'(o)$ to pick the SKUs of order $o$ as follow. For a each location $l \in \mathcal{L}$, $a_{or'(o)}^l = \sum_{s\in \mathcal{S}(o)} \xi_{ls}'$. The variable $\rho'_{or'(o)}$ associated with route $r'(o)$ is equals to 1, all other $\rho'$ variables for order $o$ are equals to 0.
\end{itemize}

It is clear that $X'$ is integral, and a valid solution. Then we prove that $X$ and $X'$ have the same cost. More precisely, for an order $o\in\mathcal{O}$, we prove that every route $r \in \mathcal{R}_o$ that is active in $X$ (i.e., $\rho_{or} > 0$), stops in the same locations than $r'(o)$ (potentially the number of stops in the locations could be different). This is equivalent to say that, for all $l \in \mathcal{L}$, $b_{or}^l = b_{or'(o)}^l$ ($b_{or}^l = 1$ if route $r \in \mathcal{R}_o$ stops in $l \in \mathcal{L}$, 0 otherwise). Then since the sum of the $\rho$ variables for an order is equal to one, this is enough to prove that $X$ and $X'$ have the same cost.\\

Let $l \in \mathcal{L}$ be a location. If $l$ is not included in $Supp_o(\xi)$ then, according to lemma 1, $b_{or}^l = b_{or'(o)}^l = 0$. If $l \in Supp_o(\xi)$, then there exists an SKU $s \in \mathcal{S}(o)$ such that $\xi_{ls} = 1$ (because the $\xi$ variables are integral). In that case, by construction, $b_{or'(o)}^l = 1$. We suppose, by absurd, that there exists an active route $r^0 \in \mathcal{R}_o$ of $X$ such that is not stopping in $l$, i.e. $b_{or^0}^l = 0$ and $\rho_{or^0} > 0$. Then according to constraint~\eqref{eq:rmp:03} we have:

$$1 = \xi_{ls} \leq \sum_{r \in \mathcal{R}_o} b_l^r\rho_{or} = \sum_{r \in \mathcal{R}_o\backslash \{r^0\}} b_l^r\rho_{or} \leq \sum_{r \in \mathcal{R}_o\backslash \{r^0\}} \rho_{or} = 1 - \rho_{or^0} < 1$$

Which is absurd and concludes the proof.
\begin{flushright}
$\square$
\end{flushright}

% end of proof
\section{Pricing problem with the largest gap routing policy}\label{appendix:largest_pricing}

The midpoint policy can be seen as the largest gap policy where the maximum turning point of a route in an aisle is fixed to be its midpoint and cannot be chosen as the position that minimizes the walking distance in it.
%The largest gap policy is similar to the midpoint, but the "middle point" of an aisle is not fixed, instead it is set at the position that minimizes the walking distance in this aisle. 
Therefore to consider the largest gap policy, additional parameters on top of those introduced by the midpoint policy are required. Note that a return policy is applied if a single aisle is visited, to avoid the creation of absurd routes. To enforce the policy, the following information is needed:
\begin{itemize}
    \item First aisle: the first aisle visited by the route. It is updated similarly to the midpoint routing policy (see Section~\ref{sec:pricing_other_policies}).    
    \item Last aisle: the last aisle visited by the route. It is updated similarly to the midpoint routing policy (see Section~\ref{sec:pricing_other_policies}).  
    \item Last locations from top: for each aisle, it stores the last location visited by the path coming from the top. When entering an aisle by the top, this parameter is updated at each new extension in the aisle. This entry may be undefined for some aisles.
    \item Largest gaps: stores the current largest gap in each aisle, i.e. the gap between two consecutive locations. Before reaching the last aisle, it is computed as the distance between the current location of a node and the bottom cross aisle. When the last aisle is chosen, this is computed as the minimum between the previous largest gap, and the new gap created by the extension, i.e. between the current location and the last one visited from the top (which we registered in the label). This entry may be undefined for some aisles.
    \item Minimum largest gaps: stores the minimum largest gap on each aisle, which corresponds to the minimal possible value for the largest gap to remain where it is supposed to be. Indeed, the resource extensions are computed with the assumption that the locations are visited by the route in the same order as the partial path extension. If the previously set largest gap in an aisle becomes smaller than the computed minimum, its position changes, and the current partial path becomes invalid, i.e. it does not respect the largest gap policy anymore. Thus, If the current largest gap becomes smaller than this value, the current label is discarded. This entry may be undefined for some aisles.
\end{itemize}

\begin{figure}[h!]
    \centering
    \includegraphics{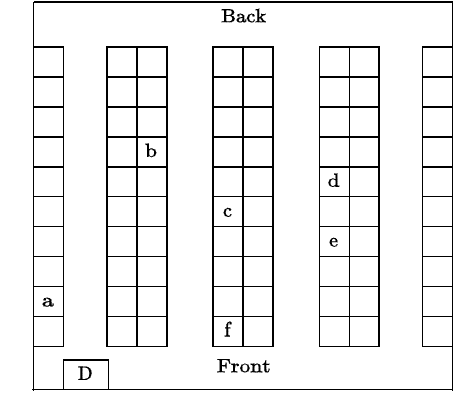}
    \caption{Example of resource extension for largest gap policy}
    \label{fig:pricing_largest}
\end{figure}

To illustrate these notions, we propose a minimal example based on Figure~\ref{fig:pricing_largest}. A step-by-step label extension is presented in Table~\ref{tab:labelextension}. Consider the route $(D,a,b,c,d,e,f,D)$. We use the symbol $\infty$ when the corresponding value is not defined, and suppose that the distance between two consecutive bays, and the distance between two consecutive aisles, are both equal to 1. The status of the resources in each step is defined as follows:

\begin{table}[h!]
    \centering
    \begin{tabular}{|c|c|c|c|c|c|c|}
    \hline
        Node & Distance & First aisle &  Last aisle & Last loc. from top & Largest gaps & Minimum largest gaps \\ 
        \hline
        $D $ & 0   & $\infty$ &  $\infty$ &$(\infty, \infty, \infty, \infty)$ & $(11, 11, 11, 11)$ & $(\infty,\infty,\infty,\infty)$\\
        $a$ & 2   & 1 &  $\infty$ &$(2, \infty, \infty, \infty)$ & $(9,11,11,11)$ & $(\infty,\infty,\infty,\infty)$\\
        $b$ & 16 & 1 & $\infty$ & $(2, 7, \infty, \infty)$ & $(9,7,11,11)$& $(\infty,4,\infty,\infty)$\\
        $c$ & 18 & 1 & $\infty$ & $(2, 5, \infty, \infty)$ & $(9,5,11,11)$ & $(\infty,4,\infty,\infty)$\\
        $d$ & 30 & 1 & $\infty$ & $(2, 5, 6, \infty)$ & $(9,5,6,11)$ & $(\infty,4,5,\infty)$\\
        $e$ & 32 & 1 & 3 & $(2, 5, 4, \infty)$ & $(9,5,4,11)$ & $(\infty,4,5,\infty)$\\
        $f$ & 38 & 1 & 3 & $(2, 5, 4, \infty)$ & $(9,4,4,11)$ & $(\infty,4,5,\infty)$\\
        $D$ & 40 & 1 & 3 & $(2, 5, 4, \infty)$ & $(9,4,4,11)$ & $(\infty,4,5,\infty)$\\
    \hline
    \end{tabular}
    \caption{Step-by-step label extension for the largest gap routing policy}
    \label{tab:labelextension}    
\end{table}

% \textbf{Extension $D \rightarrow a$}:
% \begin{itemize}
%     \item First aisle: 1. 
%     \item Last aisle: $\infty$.
%     \item Last locations from top: $(2, \infty, \infty, \infty)$
%     \item Largest gaps: $(9,11,11,11)$
%     \item Minimum largest gaps: $(\infty,\infty,\infty,\infty)$
% \end{itemize}

% \textbf{Extension $a \rightarrow b$}:
% \begin{itemize}
%     \item First aisle: 1. Last aisle: undefined.
%     \item Last location from top: [2, 7, undef, undef]
%     \item Largest gap: [9,7,11,11]
%     \item Minimum largest gap: [undef,4,undef,undef]
% \end{itemize}

% \textbf{Extension $b \rightarrow c$}:
% \begin{itemize}
%     \item First aisle: 1. Last aisle: undefined.
%     \item Last location from top: [2, 5, undef, undef]
%     \item Largest gap: [9,5,11,11]
%     \item Minimum largest gap: [undef,4,undef,undef]
% \end{itemize}

% \textbf{Extension $c \rightarrow d$}:
% \begin{itemize}
%     \item First aisle: 1. Last aisle: undefined.
%     \item Last location from top: [2, 5, 6, undef]
%     \item Largest gap: [9,5,6,11]
%     \item Minimum largest gap: [undef,4,5,undef]
% \end{itemize}

% \textbf{Extension $d \rightarrow e$}:
% \begin{itemize}
%     \item First aisle: 1. Last aisle: 3.
%     \item Last location from top: [2, 5, 4, undef]
%     \item Largest gap: [9,5,4,11]
%     \item Minimum largest gap: [undef,4,5,undef]
% \end{itemize}

Note that when extending the label from location $d$ to location $e$, the largest gap for the third aisle became smaller than the minimum largest gap, meaning that the \lq\lq middle point" of the aisle would be above location $d$. We want to avoid this situation, as the label has been extended up to $e$ with the supposition that the picker is coming from the top. The third aisle is therefore labeled as the \textit{last aisle}, making the wrong placement of the largest gap irrelevant.

% \textbf{Extension $e \rightarrow f$}:
% \begin{itemize}
%     \item First aisle: 1. Last aisle: 3.
%     \item Last location from top: [2, 5, 4, undef]
%     \item Largest gap: [9,4,4,11]
%     \item Minimum largest gap: [undef,4,5,undef]
% \end{itemize}

Note that extending the label from location $e$ to location $f$ is feasible, but that would not be the case if $f$ was farther from the front aisle: This would lead to a largest gap (between $c$ and $f$) smaller than the minimum largest gap (which is the gap between $b$ and the back cross aisle).
\section{Multi-commodity flow formulation}\label{appendix:compact_mcf}

The notations used in this formulation correspond to the one used in the paper, and are summarized in Table~\ref{tab:notations}. Subtour elimination constraints are modeled using the flow variable $g$ such that, for order $o \in \mathcal{O}$, $g_{ij}^{os}$ represents the flow of commodity $s \in \mathcal{S}(o)$ along the arc $(i,j) \in \mathcal{V}^0$.

\begin{align}
    &\textbf{(MCF)} &\min\, & \sum_{o \in \mathcal{O}} f(\mathbf{\underline{x^o}}) & & \label{eq:c1:obj:mcf}\\
    & &    s.t.\, & \sum_{s \in \mathcal{S}} \xi_{ls} \leq K_l   & \forall l \in \mathcal{L} & \label{eq:c1:01:mcf} \\
    & &    & \sum_{l \in \mathcal{L}} \xi_{ls} = 1  & \forall s \in \mathcal{S} & \label{eq:c1:02:mcf}\\
    & &    & \xi_{ls} = 1 & \forall (s,l) \in \mathcal{F} & \label{eq:c1:022:mcf}\\
    & &    & \sum_{v_i \in \mathcal{V}} x_{i0}^o = \sum_{v_j \in \mathcal{V}} x_{0j}^o = 1  & \forall o \in \mathcal{O} & \label{eq:c1:03:mcf}\\
    & &    & \sum_{v_j \in \mathcal{V}^0} x_{ij}^o = \sum_{v_j \in \mathcal{V}^0} x_{ji}^o & \forall v_i \in \mathcal{V}, o \in \mathcal{O} & \label{eq:c1:04:mcf} \\
    %& &    & u^o_j \geq u^o_i + 1 - |\mathcal{S}(o)|\, (1-x_{ij}^o) & \forall o \in \mathcal{O}, v_i,v_j \in \mathcal{V}& \label{eq:c1:05}\\
    & &    &\sum_{v_i \in \mathcal{V}^0} \sum_{v_j \in \mathcal{V}(l)} (g^{os}_{ij} - g^{os}_{ji}) = \xi_{ls} & \forall o \in \mathcal{O}, s \in \mathcal{S}(o), l \in \mathcal{L} & \label{eq:c1:05bis:mcf} \\
    & &    & \sum_{v_i \in \mathcal{V}^0} \sum_{v_j \in \mathcal{V}} x_{ij}^o = |\mathcal{S}(o)| & \forall o \in \mathcal{O} & \label{eq:c1:06:mcf}\\
    & &    & \sum_{v_i \in \mathcal{V}^0} \sum_{v_j \in \mathcal{V}(l)} x_{ij}^o \geq \sum_{s \in \mathcal{S}(o)} \xi_{ls} & \forall l \in \mathcal{L}, o \in \mathcal{O} & \label{eq:c1:07:mcf} \\
    & &    & 0 \leq g^{os}_{ij} \leq x^o_{ij} & \forall o \in \mathcal{O}, s \in \mathcal{S}(o), i,j \in \mathcal{V}^0 & \label{eq:c1:10bis:mcf} \\
    & &    & x_{ij}^o \in \{0,1\} & \forall o \in \mathcal{O}, l,h \in \mathcal{V}^0 & \label{eq:c1:08:mcf}\\
    & &    & \xi_{ls} \in \{0,1\} & \forall l \in \mathcal{L}, s \in \mathcal{S}&\label{eq:c1:09:mcf}
    %& &    & 0 \leq u^o_i \leq |\mathcal{S}(o)| - 1 & \forall o \in \mathcal{O}, v_i \in \mathcal{V} & \label{eq:c1:10}
\end{align}

This formulation is very similar to formulation (C) introduced in Section~\ref{sec:problem_description}, with only constraints~\eqref{eq:c1:05bis:mcf} and~\eqref{eq:c1:10bis:mcf} being different. Constraints~\eqref{eq:c1:05bis:mcf} ensure that the difference between the in-flow and out-flow of commodity $s \in \mathcal{S}(o)$ is equal to one in location $l \in \mathcal{L}$ if SKU $s$ is assigned to $l$, and 0 otherwise. Constraints~\eqref{eq:c1:10bis:mcf} define the domain of variables $g$, and ensure that the flow of the commodities are set to zero on the arcs that are not used in the routes.

\section{Compact formulation for return routing policy}\label{appendix:compact_return}

Appendices~\ref{appendix:compact_return}-\ref{appendix:compact_largest} present compact formulations for the SLAPRP with classical routing policies. These formulations are heavily inspired by the work of \cite{silva_integrating_2020}. The main difference is that we linearize the conditional constraints (i.e. if-then-else constraints), leading to a slight performance improvement when solved with a commercial solver (see Section~\ref{sec:results:silva}).

\begin{table}[h]
\centering
\caption{Mathematical notations for return policy}
\begin{tabular}{ll}
\hline
\multicolumn{2}{l}{Sets}       \\ \hline
$\mathcal{A} = \{1,\dots \overline{a}\}$ & Set of aisles  \\
$\mathcal{B} = \{1,\dots \overline{b}\}$ & Set of bays \\
$\mathcal{S}$ & Set of SKUs \\
$\mathcal{O}$ & Set of orders \\
$\mathcal{S}(o)$ & Set of SKUs of order $o \in \mathcal{O}$ \\
\hline
\multicolumn{2}{l}{Parameters} \\ \hline
$D$ & Distance between two consecutive aisles \\
$d$ & Distance between two consecutive bays \\
\hline
\multicolumn{2}{l}{Variables}  \\ \hline
$\xi_{ab}^s$   &  1 if SKU $s \in \mathcal{S}$ is stored in aisle $a \in \mathcal{A}$ and bay $b \in \mathcal{B}$, 0 otherwise\\
$f_{oa}$ & Farthest bay visited in aisle $a \in \mathcal{A}$ when picking order $o \in \mathcal{O}$ \\
$z_{oa}$ & 1 if aisle $a \in \mathcal{A}$ is visited when picking order $o \in \mathcal{O}$, 0 otherwise \\
$v_o$ & Farthest aisle visited when picking order $o \in \mathcal{O}$\\
\hline
\end{tabular}
\end{table}

\begin{align}
     Min \; & 2 \,\sum_{o \in \mathcal{O}} \left(D(v_o-1) + d\sum_{a \in \mathcal{A}}f_{oa}\right) & & \\
     s.t.\; & \sum_{s \in \mathcal{S}} \xi_{ab}^s \leq 2 & \forall a \in \mathcal{A}, b \in \mathcal{B} & \label{eq:return:01} \\
     & \sum_{a \in \mathcal{A}} \sum_{b \in \mathcal{B}} \xi_{ab}^s = 1 & \forall s \in \mathcal{S} & \label{eq:return:02}\\
     & f_{oa} \geq \sum_{b \in \mathcal{B}} b \xi_{ab}^s & \forall o \in \mathcal{O}, a \in \mathcal{A}, s \in \mathcal{S}(o) & \label{eq:return:03}\\
     & f_{oa} \leq \overline{b}z_{oa} & \forall o \in \mathcal{O}, a \in \mathcal{A} & \label{eq:return:04}\\
     & z_{oa} \rightarrow (v_{o} \geq a) & \forall o \in \mathcal{O}, a \in \mathcal{A} & \label{eq:return:05}\\
     & \xi_{ab}^s \in \{0,1\} & \forall a \in \mathcal{A}, b \in \mathcal{B}, s \in \mathcal{S} & \label{eq:return:06}\\
     & 0 \leq f_{oa} \leq \overline{b} & \forall o \in \mathcal{O}, a \in \mathcal{A} & \label{eq:return:07}\\
     & z_{oa} \in \{0,1\} & \forall o \in \mathcal{O}, a \in \mathcal{A} & \label{eq:return:08} \\
     & 1 \leq v_o \leq \overline{a} & \forall o \in \mathcal{O} & \label{eq:return:09}
\end{align}

The objective function minimizes the total walking distance, computed as twice the horizontal distance to the farthest aisle visited (first term), plus twice the vertical distance to the farthest pick in each aisle (second term). Constraints~\eqref{eq:return:01} and \eqref{eq:return:02} are the assignment constraints. Constraints~\eqref{eq:return:03} ensure that $f$ is correctly defined. Constraints~\eqref{eq:return:04} ensure that $z_{oa}$ is set to one if the aisle $a \in \mathcal{A}$ is crossed when picking order $o \in \mathcal{O}$. Constraints~\eqref{eq:return:05} ensure that $v_o$ is correctly defined, i.e. if the route visits aisle $a \in \mathcal{A}$, then $v_o$ should be greater than $a$. Constraints~\eqref{eq:return:06}-\eqref{eq:return:09} define the domains of the variables.

\section{Compact formulation for S-shape routing policy}\label{appendix:compact_sshape}

\begin{table}[h]
\centering
\caption{Mathematical notations for S-shape policy}
\begin{tabular}{ll}
\hline
% \multicolumn{2}{l}{Sets}       \\ \hline
% $\mathcal{A} = \{1,\dots \overline{a}\}$ & Set of aisles  \\
% $\mathcal{B} = \{1,\dots \overline{b}\}$ & Set of bays \\
% $\mathcal{S}$ & Set of SKUs \\
% $\mathcal{O}$ & Set of orders \\
% $\mathcal{S}(o)$ & Set of SKUs of order $o \in \mathcal{O}$ \\
% \hline
% \multicolumn{2}{l}{Parameters} \\ \hline
% $D$ & Distance between two consecutive aisles \\
% $d$ & Distance between two consecutive bays \\
% \hline
\multicolumn{2}{l}{Variables}  \\ \hline
$\xi_{ab}^s$   &  1 if SKU $s \in \mathcal{S}$ is stored in aisle $a \in \mathcal{A}$ and bay $b \in \mathcal{B}$, 0 otherwise\\
$f_{oa}$ & Farthest bay visited in aisle $a \in \mathcal{A}$ when picking order $o \in \mathcal{O}$ \\
$z_{oa}$ & 1 if aisle $a \in \mathcal{A}$ is visited when picking order $o \in \mathcal{O}$, 0 otherwise \\
$v_{oa}$ & 1 if aisle $a \in \mathcal{A}$ is the farthest aisle visited when picking order $o \in \mathcal{O}$, 0 otherwise\\
$s_o$ & 1 if the number of visited aisles for order $o \in \mathcal{O}$ is odd, 0 otherwise \\
$k_o$ & Auxiliary variable to compute $s_o$\\
$\alpha_{oa}$ & Linearization variable, with $\alpha_{oa} = s_of_{oa}v_{oa}$ \\
\hline
\end{tabular}
\end{table}
\vspace*{-1cm}
\begin{align}
     Min \; & \, 2D\sum_{o \in \mathcal{O}} \sum_{a \in \mathcal{A}} (a-1)v_{oa} + d\sum_{o \in \mathcal{O}} \sum_{a \in \mathcal{A}} (( \overline{b} + 1)z_{oa} + 2\alpha_{oa}) & & \\
     \nonumber & - 2d (\overline{b} + 1) \sum_{o \in \mathcal{O}} s_o & & \\
     s.t.\; & \sum_{s \in \mathcal{S}} \xi_{ab}^s \leq 2 & \forall a \in \mathcal{A}, b \in \mathcal{B} & \label{eq:sshape:01} \\
     & \sum_{a \in \mathcal{A}} \sum_{b \in \mathcal{B}} \xi_{ab}^s = 1 & \forall s \in \mathcal{S} & \label{eq:sshape:02}\\
     & f_{oa} \geq \sum_{b \in \mathcal{B}} b \xi_{ab}^s & \forall o \in \mathcal{O}, a \in \mathcal{A}, s \in \mathcal{S}(o) &\label{eq:sshape:03} \\
     & f_{oa} \leq \overline{b}z_{oa} & \forall o \in \mathcal{O}, a \in \mathcal{A} &\label{eq:sshape:04} \\
     & !z_{oa} \rightarrow (v_{oa} \leq 0) & \forall o \in \mathcal{O}, a \in \mathcal{A} &\label{eq:sshape:05} \\
     & z_{oa} \rightarrow ( \sum_{c=1}^{a-1} v_{oc} <= 0 ) & \forall o \in \mathcal{O}, a \in \mathcal{A} &\label{eq:sshape:06} \\
     & \sum_{a \in \mathcal{A}} v_{oa} = 1 & \forall o \in \mathcal{O} &\label{eq:sshape:07} \\
     & \sum_{a \in \mathcal{A}} z_{oa} = 2k_o + s_o & \forall o \in \mathcal{O} &\label{eq:sshape:08} \\
     & \alpha_{oa} \leq \overline{b} s_o & \forall o \in \mathcal{O}, a \in \mathcal{A} &\label{eq:sshape:09} \\
     & \alpha_{oa} \leq f_{oa} & \forall o \in \mathcal{O}, a \in \mathcal{A} &\label{eq:sshape:10} \\
     & \alpha_{oa} \leq \overline{b} v_{oa} & \forall o \in \mathcal{O}, a \in \mathcal{A} &\label{eq:sshape:11} \\
     & \alpha_{oa} \geq f_{oa} -  \overline{b}(2 - s_o - v_{oa}) & \forall o \in \mathcal{O}, a \in \mathcal{A} &\label{eq:sshape:12} \\
     & \xi_{ab}^s \in \{0,1\} & \forall a \in \mathcal{A}, b \in \mathcal{B}, s \in \mathcal{S} &\label{eq:sshape:13} \\
     & 0 \leq f_{oa} \leq \overline{b} & \forall o \in \mathcal{O}, a \in \mathcal{A} &\label{eq:sshape:14} \\
     & z_{oa},\, v_{oa},\, s_o \in \{0,1\} & \forall o \in \mathcal{O}, a \in \mathcal{A} &\label{eq:sshape:15} \\
     & k_o \in \mathbb{N} \cap [0,\frac{\overline{a}}{2}] & \forall o \in \mathcal{O} &\label{eq:sshape:18} \\
     & \alpha_{oa} \geq 0 & \forall o \in \mathcal{O}, a \in \mathcal{A} &\label{eq:sshape:19} 
\end{align}

The first term of the objective function is the horizontal distance, up to the farthest aisle and back. The second term $(\overline{b}+1)z_{oa}$ is the distance in aisle $a \in \mathcal{A}$ for order $o \in \mathcal{O}$ when the aisle is entirely crossed. The third term $2\alpha_{oa}$ is the distance in the last aisle traversed by order $o \in \mathcal{O}$, computed as a return. This term is equal to 0 if the last aisle is crossed entirely (i.e. $s_o = 0$). The last term corrects the value of the objective function if the last aisle is not crossed entirely. Constraints~\eqref{eq:sshape:01}-\eqref{eq:sshape:04} are similar to the ones defined for the return routing in Appendix~\ref{appendix:compact_return}. Constraints~\eqref{eq:sshape:05}-\eqref{eq:sshape:07} ensure that $v_{oa}$ is computed correctly. Constraints~\eqref{eq:sshape:08} compute $s_o$ as the remainder of $\frac{\sum_{a \in \mathcal{A}}z_{oa}}{2}$. Constraints~\eqref{eq:sshape:09}-\eqref{eq:sshape:12} compute $\alpha_{oa}$ as $s_of_{oa}v_{oa}$. Constraints~\eqref{eq:sshape:13}-\eqref{eq:sshape:19} define the domains of the variables.
\section{Compact formulation for midpoint routing policy}\label{appendix:compact_midpoint}

\begin{table}[h]
\centering
\caption{Mathematical notations for midpoint policy}
\begin{tabular}{ll}
\hline
% \multicolumn{2}{l}{Sets}       \\ \hline
% $\mathcal{A} = \{1,\dots \overline{a}\}$ & Set of aisles  \\
% $\mathcal{B} = \{1,\dots \overline{b}\}$ & Set of bays \\
% $\mathcal{S}$ & Set of SKUs \\
% $\mathcal{O}$ & Set of orders \\
% $\mathcal{S}(o)$ & Set of SKUs of order $o \in \mathcal{O}$ \\
% \hline
% \multicolumn{2}{l}{Parameters} \\ \hline
% $D$ & Distance between two consecutive aisles \\
% $d$ & Distance between two consecutive bays \\
% $mp$ & Middle point of each aisle \\
% \hline
\multicolumn{2}{l}{Variables}  \\ \hline
$\xi_{ab}^s$   &  1 if SKU $s \in \mathcal{S}$ is stored in aisle $a \in \mathcal{A}$ and bay $b \in \mathcal{B}$, 0 otherwise\\
$f_{oa}$ & Farthest bay visited in aisle $a \in \mathcal{A}$ when picking order $o \in \mathcal{O}$ \\
$f_{oa}^-$ & Farthest bay visited in aisle $a \in \mathcal{A}$ when picking order $o \in \mathcal{O}$, below the midpoint\\
$f_{oa}^+$ & Farthest bay visited in aisle $a \in \mathcal{A}$, from the back cross aisle and above midpoint, \\
& when picking order $o \in \mathcal{O}$\\
$z_{oa}$ & 1 if aisle $a \in \mathcal{A}$ is visited when picking order $o \in \mathcal{O}$, 0 otherwise \\
$v_o$ & Farthest aisle visited when picking order $o \in \mathcal{O}$\\
$v_{oa}$ & 1 if $a \in \mathcal{A}$ is the farthest aisle visited when picking order $o \in \mathcal{O}$, 0 otherwise\\
$u_o$ & Closest aisle visited when picking order $o \in \mathcal{O}$ \\
$u_{oa}$ & 1 if $a \in \mathcal{A}$ is the closest aisle visited when picking order $o \in \mathcal{O}$, 0 otherwise\\
$s_o$ & 1 if at least two aisles are visited when picking order $o \in \mathcal{O}$, 0 otherwise\\
$w_{oa}$ & 1 if aisle $a \in \mathcal{A}$ is (strictly) between the first and last aisles visited\\
& when picking order $o \in \mathcal{O}$, 0 otherwise\\
$\alpha_{oa}$ & Auxiliary variable used to define $u_o$\\
$\beta_{oa}$ & Linearization variable, with $\beta_{oa} = s_ow_{oa}(f^-_{oa} + f^+_{oa})$\\
$\gamma_{oa}$ & Linearization variable, with $\gamma_{oa} = f_{oa}v_{oa}(1-s_o)$\\
\hline
\end{tabular}
\end{table}

\begin{align}
     Min \; & 2D\sum_{o \in \mathcal{O}} (v_o - 1) + 2d(\overline{b}+1)\sum_{o \in \mathcal{O}} s_o + 2d\sum_{o \in \mathcal{O}}\sum_{a \in \mathcal{A}} (\beta_{oa} + \gamma_{oa}) & & \\
     s.t.\; & \sum_{s \in \mathcal{S}} \xi_{ab}^s \leq 2 & \forall a \in \mathcal{A}, b \in \mathcal{B} & \label{eq:midpoint:01}\\
     & \sum_{a \in \mathcal{A}} \sum_{b \in \mathcal{B}} \xi_{ab}^s = 1 & \forall s \in \mathcal{S} & \label{eq:midpoint:02}\\
     & f_{oa} \geq \sum_{b \in \mathcal{B}} b \xi_{ab}^s & \forall o \in \mathcal{O}, a \in \mathcal{A}, s \in \mathcal{S}(o) & \label{eq:midpoint:03}\\
     & f_{oa} \leq \overline{b}z_{oa} & \forall o \in \mathcal{O}, a \in \mathcal{A} & \label{eq:midpoint:04}\\
     & z_{oa} \rightarrow (v_{o} \geq a) & \forall o \in \mathcal{O}, a \in \mathcal{A} & \label{eq:midpoint:05}\\
     & v_o = \sum_{a \in \mathcal{A}} av_{oa} & \forall o \in \mathcal{O} & \label{eq:midpoint:06}\\
     & \sum_{a \in \mathcal{A}} v_{oa} = 1 & \forall o \in \mathcal{O} & \label{eq:midpoint:07}\\
     & z_{oa} \rightarrow ( u_o \leq a) & \forall o \in \mathcal{O}, a \in \mathcal{A} & \label{eq:midpoint:08}\\
     & \alpha_{oa} \leq \sum_{c=1}^{a-1} z_{oc} & \forall o \in \mathcal{O}, a \in \mathcal{A} & \label{eq:midpoint:09}\\
     & ! \alpha_{oa} \rightarrow (u_o \geq a) & \forall o \in \mathcal{O}, a \in \mathcal{A} & \label{eq:midpoint:10}\\
     & u_o = \sum_{a \in \mathcal{A}} au_{oa} & \forall o \in \mathcal{O} & \label{eq:midpoint:11}\\
     & \sum_{a \in \mathcal{A}} u_{oa} = 1 & \forall o \in \mathcal{O} & \label{eq:midpoint:12}\\
     & w_{oa} \leq \sum_{c=1}^{a-1} u_{oc} & \forall o \in \mathcal{O}, a \in \mathcal{A} & \label{eq:midpoint:13}\\
     & w_{oa} \leq \sum_{c=a+1}^{\overline{a}} v_{oc} & \forall o \in \mathcal{O}, a \in \mathcal{A} & \label{eq:midpoint:14}\\
     & w_{oa} \geq \sum_{c=1}^{a-1} u_{oc} + \sum_{c=a+1}^{\overline{a}} v_{oc} - 1 & \forall o \in \mathcal{O}, a \in \mathcal{A} & \label{eq:midpoint:15}\\
     & s_o \leq v_o - u_o & \forall o \in \mathcal{O} & \label{eq:midpoint:16}\\
     & v_{oa} \rightarrow (s_o \geq 1 - u_{oa}) & \forall o \in \mathcal{O}, a \in \mathcal{A} & \label{eq:midpoint:17}\\
     & f^-_{oa} \geq \sum_{b \in \mathcal{B}^-} b \xi_{ab}^s & \forall o \in \mathcal{O}, a \in \mathcal{A}, s \in \mathcal{S}(o) & \label{eq:midpoint:18}\\
     & f^+_{oa} \geq \sum_{b \in \mathcal{B}^+} (\overline{b} + 1 - b) \xi_{ab}^s & \forall o \in \mathcal{O}, a \in \mathcal{A}, s \in \mathcal{S}(o) & \label{eq:midpoint:19}\\
     & \beta_{oa} \leq (\overline{b} + 1)s_o & \forall o \in \mathcal{O}, a \in \mathcal{A} & \label{eq:midpoint:20}\\
     & \beta_{oa} \leq f^-_{oa} + f^+_{oa} & \forall o \in \mathcal{O}, a \in \mathcal{A} & \label{eq:midpoint:21}\\
     & \beta_{oa} \leq (\overline{b} + 1) w_{oa} & \forall o \in \mathcal{O}, a \in \mathcal{A} & \label{eq:midpoint:22}\\
     & \beta_{oa} \geq (f^-_{oa} + f^+_{oa}) - (\overline{b} + 1)(2 - s_o - w_{oa}) & \forall o \in \mathcal{O}, a \in \mathcal{A} & \label{eq:midpoint:23}\\
     & \gamma_{oa} \leq \overline{b}(1-s_o) & \forall o \in \mathcal{O}, a \in \mathcal{A} & \label{eq:midpoint:24}\\
     & \gamma_{oa} \leq f_{oa} & \forall o \in \mathcal{O}, a \in \mathcal{A} & \label{eq:midpoint:25}\\
     & \gamma_{oa} \leq \overline{b} v_{oa} & \forall o \in \mathcal{O}, a \in \mathcal{A} & \label{eq:midpoint:26}\\
     & \gamma_{oa} \geq f_{oa} - \overline{b} (2 - (1 - s_o) - v_{oa}) & \forall o \in \mathcal{O}, a \in \mathcal{A} & \label{eq:midpoint:27}\\
     & \xi_{ab}^s \in \{0,1\} & \forall a \in \mathcal{A}, b \in \mathcal{B}, s \in \mathcal{S} & \label{eq:midpoint:28}\\
     & 0 \leq f_{oa} \leq \overline{b} & \forall o \in \mathcal{O}, a \in \mathcal{A} & \label{eq:midpoint:29}\\
     %& z_{oa} \in \{0,1\} & \forall o \in \mathcal{O}, a \in \mathcal{A} & \label{eq:midpoint:30}\\
     & 1 \leq v_o \leq \overline{a} & \forall o \in \mathcal{O} & \label{eq:midpoint:31}\\
     %& v_{oa} \in \{0,1\} & \forall o \in \mathcal{O}, a \in \mathcal{A} & \label{eq:midpoint:32}\\
     %& u_o \geq 0 & \forall o \in \mathcal{O} & \label{eq:midpoint:33}\\
     %& u_{oa} \in \{0,1\} & \forall o \in \mathcal{O}, a \in \mathcal{A} & \label{eq:midpoint:34}\\
     %& s_o \in \{0,1\} & \forall o \in \mathcal{O} & \label{eq:midpoint:35}\\
     %& w_{oa} \in \{0,1\} & \forall o \in \mathcal{O}, a \in \mathcal{A} & \label{eq:midpoint:36}\\
     & 0 \leq f^-_{oa} \leq mp & \forall o \in \mathcal{O}, a \in \mathcal{A} & \label{eq:midpoint:37}\\
     & 0 \leq f^+_{oa} \leq \overline{b} - mp & \forall o \in \mathcal{O}, a \in \mathcal{A} & \label{eq:midpoint:38}\\
     & u_o,\, s_o,\, u_{oa},\, v_{oa},\, w_{oa},\, z_{oa},\, \alpha_{oa}, \in \{0,1\} & \forall o \in \mathcal{O}, a \in \mathcal{A} & \label{eq:midpoint:39}\\
     & \beta_{oa},\, \gamma_{oa} \geq 0 & \forall o \in \mathcal{O}, a \in \mathcal{A} & \label{eq:midpoint:41}
\end{align}

The first term of the objective function is the horizontal distance, up to the farthest aisle and back. The second term is the vertical distance to cross entirely the first and last aisles, if at least two aisles are visited (represented by $s_o$). The last term is composed of two parts: the one related to variable $\beta_{oa}$ is the vertical distance in the visited aisles that are neither the first, nor the last one visited. The term related to $\gamma_{oa}$ is the vertical distance in case of a single aisle is visited (in this case a return policy is applied). Constraints~\eqref{eq:midpoint:01}-\eqref{eq:midpoint:04} are identical to the return policy. Constraints~\eqref{eq:midpoint:05}-\eqref{eq:midpoint:07} ensure that the last aisle $v_o$, and its binary equivalent $v_{oa}$, are correctly defined. Constraints~\eqref{eq:midpoint:08}-\eqref{eq:midpoint:12} ensure that the first aisle $u_o$, and its binary equivalent $u_{oa}$, are correctly defined. Constraints~\eqref{eq:midpoint:13}-\eqref{eq:midpoint:15} ensure that $w_{oa}$ is correctly defined, valuing 1 for an aisle that is strictly between the first and the last ones, 0 otherwise. Constraints~\eqref{eq:midpoint:16}-\eqref{eq:midpoint:17} define $s_o$. Constraints~\eqref{eq:midpoint:18}-\eqref{eq:midpoint:19} compute the distances walked below the midpoint ($f^-_{oa}$) and above the midpoint ($f^+_{oa}$). Constraints~\eqref{eq:midpoint:20}-\eqref{eq:midpoint:23} compute $\beta_{oa} = s_ow_{oa}(f^-_{oa} + f^+_{oa})$. Constraints~\eqref{eq:midpoint:24}-\eqref{eq:midpoint:27} compute $\gamma_{oa} = f_{oa}v_{oa}(1-s_o)$. Constraints~\eqref{eq:midpoint:28}-\eqref{eq:midpoint:41} define the domains of the variables.

\section{Compact formulation for largest gap routing policy}\label{appendix:compact_largest}

\begin{table}[h]
\centering
\caption{Mathematical notations for largest gap policy}
\begin{tabular}{ll}
\hline
% \multicolumn{2}{l}{Sets}       \\ \hline
% $\mathcal{A} = \{1,\dots \overline{a}\}$ & Set of aisles  \\
% $\mathcal{B} = \{1,\dots \overline{b}\}$ & Set of bays \\
% $\mathcal{B}^0 = \mathcal{B} \cup \{0\}$ & Set of bays with a dummy one \\
% $\mathcal{S}$ & Set of SKUs \\
% $\mathcal{O}$ & Set of orders \\
% $\mathcal{S}(o)$ & Set of SKUs of order $o \in \mathcal{O}$ \\
% \hline
% \multicolumn{2}{l}{Parameters} \\ \hline
% $D$ & Distance between two consecutive aisles \\
% $d$ & Distance between two consecutive bays \\
% \hline
\multicolumn{2}{l}{Variables}  \\ \hline
$\xi_{ab}^s$   &  1 if SKU $s \in \mathcal{S}$ is stored in aisle $a \in \mathcal{A}$ and bay $b \in \mathcal{B}$, 0 otherwise\\
$f_{oa}$ & Farthest bay visited in aisle $a \in \mathcal{A}$ when picking order $o \in \mathcal{O}$ \\
$z_{oa}$ & 1 if aisle $a \in \mathcal{A}$ is visited when picking order $o \in \mathcal{O}$, 0 otherwise \\
$v_o$ & Farthest aisle visited when picking order $o \in \mathcal{O}$\\
$v_{oa}$ & 1 if $a \in \mathcal{A}$ is the farthest aisle visited when picking order $o \in \mathcal{O}$, 0 otherwise\\
$u_o$ & Closest aisle visited when picking order $o \in \mathcal{O}$ \\
$u_{oa}$ & 1 if $a \in \mathcal{A}$ is the closest aisle visited when picking order $o \in \mathcal{O}$, 0 otherwise\\
$s_o$ & 1 if at least two aisles are visited when picking order $o \in \mathcal{O}$, 0 otherwise\\
$w_{oa}$ & 1 if aisle $a \in \mathcal{A}$ is (strictly) between the first and last aisles visited\\
& when picking order $o \in \mathcal{O}$, 0 otherwise\\
$G_{oa}$ & Length of the largest gap in aisle $a \in \mathcal{A}$ when picking order $o \in \mathcal{O}$\\
$g_{oab}$ & Gap between a pick in bay $b\in\mathcal{B}^0$ and another pick (or the back cross aisle) in aisle $a \in \mathcal{A}$, \\
& when picking order $o \in \mathcal{O}$\\
$h_{oab}$ & 1 if $g_{oab}$ is the largest gap in aisle $a \in \mathcal{A}$ for $b \in \mathcal{B}^0$ when picking order $o \in \mathcal{O}$, 0 otherwise\\
$\alpha_{oa}$ & Auxiliary variable used to define $u_o$\\
$\beta_{oa}$ & Auxiliary variable used to define $g_{oab}$\\
$\gamma_{oa}$ & Linearization variable, with $\gamma_{oa} = s_ow_{oa}(\overline{b} + 1 - G_{oa})$\\
$\delta_{oa}$ & Linearization variable, with $\delta_{oa} = f_{oa}v_{oa}(1 - s_o)$\\
\hline
\end{tabular}
\end{table}

\begin{align}
    Min \; & 2D\sum_{o \in \mathcal{O}} (v_o - 1) + 2d(\overline{b}+1)\sum_{o \in \mathcal{O}} s_o & & \\
    \nonumber & + 2d\sum_{o \in \mathcal{O}}\sum_{a \in \mathcal{A}} (\gamma_{oa} + \delta_{oa}) & &  \\
     s.t.\; & \sum_{s \in \mathcal{S}} \xi_{ab}^s \leq 2 & \forall a \in \mathcal{A}, b \in \mathcal{B} & \label{eq:largest:01}\\
     & \sum_{a \in \mathcal{A}} \sum_{b \in \mathcal{B}} \xi_{ab}^s = 1 & \forall s \in \mathcal{S} & \label{eq:largest:02}\\
     & f_{oa} \geq \sum_{b \in \mathcal{B}} b \xi_{ab}^s & \forall o \in \mathcal{O}, a \in \mathcal{A}, s \in \mathcal{S}(o) & \label{eq:largest:03}\\
     & f_{oa} \leq \overline{b}z_{oa} & \forall o \in \mathcal{O}, a \in \mathcal{A} & \label{eq:largest:04}\\
     & z_{oa} \rightarrow (v_{o} \geq a) & \forall o \in \mathcal{O}, a \in \mathcal{A} & \label{eq:largest:05}\\
     & v_o = \sum_{a \in \mathcal{A}} av_{oa} & \forall o \in \mathcal{O} & \label{eq:largest:06}\\
     & \sum_{a \in \mathcal{A}} v_{oa} = 1 & \forall o \in \mathcal{O} & \label{eq:largest:07}\\
     & z_{oa} \rightarrow ( u_o \leq a) & \forall o \in \mathcal{O}, a \in \mathcal{A} & \label{eq:largest:08}\\
     & \alpha_{oa} \leq \sum_{c=1}^{a-1} z_{oc} & \forall o \in \mathcal{O}, a \in \mathcal{A} & \label{eq:largest:09}\\
     & ! \alpha_{oa} \rightarrow (u_o \geq a) & \forall o \in \mathcal{O}, a \in \mathcal{A} & \label{eq:largest:10}\\
     & u_o = \sum_{a \in \mathcal{A}} au_{oa} & \forall o \in \mathcal{O} & \label{eq:largest:11}\\
     & \sum_{a \in \mathcal{A}} u_{oa} = 1 & \forall o \in \mathcal{O} & \label{eq:largest:12}\\
     & w_{oa} \leq \sum_{c=1}^{a-1} u_{oc} & \forall o \in \mathcal{O}, a \in \mathcal{A} & \label{eq:largest:13}\\
     & w_{oa} \leq \sum_{c=a+1}^{\overline{a}} v_{oc} & \forall o \in \mathcal{O}, a \in \mathcal{A} & \label{eq:largest:14}\\
     & w_{oa} \geq \sum_{c=1}^{a-1} u_{oc} + \sum_{c=a+1}^{\overline{a}} v_{oc} - 1 & \forall o \in \mathcal{O}, a \in \mathcal{A} & \label{eq:largest:15}\\
     & s_o \leq v_o - u_o & \forall o \in \mathcal{O} & \label{eq:largest:16}\\
     & s_o \geq 1 - u_{oa} & \forall o \in \mathcal{O}, a \in \mathcal{A} & \label{eq:largest:17}\\
     & !z_{oa} \rightarrow (g_{oa0} = \overline{b}+1) & \forall o \in \mathcal{O}, a \in \mathcal{A} & \label{eq:largest:18}\\
     & \xi_{abs} \rightarrow (g_{oa0} \leq b) & o\in \mathcal{O}, a \in \mathcal{A}, s \in \mathcal{S}(o), b \in \mathcal{B} &\label{eq:largest:19}\\
     & \beta_{oab} \leq \sum_{s \in \mathcal{S}(o)} \xi_{abs} & \forall o \in \mathcal{O}, a \in \mathcal{A}, b \in \mathcal{B} & \label{eq:largest:20}\\
     & !\beta_{oab} \rightarrow (g_{oab} = 0) & \forall o \in \mathcal{O}, a \in \mathcal{A}, b \in \mathcal{B} &\label{eq:largest:21} \\
     & \xi_{ab's} \rightarrow (g_{oab} \leq b' - b) & \forall o \in \mathcal{O}, a \in \mathcal{A}, b,b' \in \mathcal{B}, b' > b, & \label{eq:largest:22}\\
     \nonumber & s \in \mathcal{S}(o) & \\
     & g_{oab} \leq \overline{b} + 1 - b & \forall o \in \mathcal{O}, a \in \mathcal{A}, b \in \mathcal{B} &\label{eq:largest:23} \\
     & G_{oa} \geq g_{oab} & \forall o \in \mathcal{O}, a \in \mathcal{A}, b \in \mathcal{B}^0 & \label{eq:largest:24}\\
     & G_{oa} \leq g_{oab} + (\overline{b} + 1)(1 - h_{oab}) & \forall o \in \mathcal{O}, a \in \mathcal{A}, b \in \mathcal{B}^0 & \label{eq:largest:25}\\
     & \sum_{b \in \mathcal{B}^0} h_{oab} = 1 & \forall o \in \mathcal{O}, a \in \mathcal{A} &\label{eq:largest:26} \\
     & \gamma_{oa} \leq \overline{b} (1-s_o) & \forall o \in \mathcal{O}, a \in \mathcal{A} & \label{eq:largest:27}\\
     & \gamma_{oa} \leq \overline{b} + 1 - G_{oa} & \forall o \in \mathcal{O}, a \in \mathcal{A} &\label{eq:largest:28} \\
     & \gamma_{oa} \leq (\overline{b} + 1)w_{oa} & \forall o \in \mathcal{O}, a \in \mathcal{A} &\label{eq:largest:29} \\
     & \gamma_{oa} \geq (\overline{b} + 1 - G_{oa}) - (\overline{b} + 1)(2 - s_o - w_{oa}) & \forall o \in \mathcal{O}, a \in \mathcal{A}&\label{eq:largest:30} \\
     & \delta_{oa} \leq \overline{b}(1-s_o) & \forall o \in \mathcal{O}, a \in \mathcal{A} &\label{eq:largest:31}\\
     & \delta_{oa} \leq f_{oa} & \forall o \in \mathcal{O}, a \in \mathcal{A} & \label{eq:largest:32}\\
     & \delta_{oa} \leq \overline{b}v_{oa} & \forall o \in \mathcal{O}, a \in \mathcal{A} & \label{eq:largest:33}\\
     & \delta_{oa} \geq f_{oa} - \overline{b}(2 - (1-s_o) - v_{oa}) & \forall o \in \mathcal{O}, a \in \mathcal{A} & \label{eq:largest:34}\\
     & \xi_{ab}^s \in \{0,1\} & \forall a \in \mathcal{A}, b \in \mathcal{B}, s \in \mathcal{S} &\label{eq:largest:35} \\
     & 0 \leq f_{oa} \leq \overline{b} & \forall o \in \mathcal{O}, a \in \mathcal{A} & \label{eq:largest:36}\\
     %& z_{oa} \in \{0,1\} & \forall o \in \mathcal{O}, a \in \mathcal{A} & \label{eq:largest:37}\\
     & 1 \leq v_o \leq \overline{a} & \forall o \in \mathcal{O} & \label{eq:largest:38}\\
     %& v_{oa} \in \{0,1\} & \forall o \in \mathcal{O}, a \in \mathcal{A} & \label{eq:largest:39}\\
     %& u_o \geq 0 & \forall o \in \mathcal{O} & \label{eq:largest:40}\\
     %& u_{oa} \in \{0,1\} & \forall o \in \mathcal{O}, a \in \mathcal{A} &\label{eq:largest:41} \\
     %& s_o \in \{0,1\} & \forall o \in \mathcal{O} & \label{eq:largest:42}\\
     %& w_{oa} \in \{0,1\} & \forall o \in \mathcal{O}, a \in \mathcal{A} & \label{eq:largest:43}\\
     %& G_{oa} \geq 0 & \forall o \in \mathcal{O}, a \in \mathcal{A} & \label{eq:largest:44}\\
     & g_{oab} \geq 0 & \forall o \in \mathcal{O}, a \in \mathcal{A}, b \in \mathcal{B}^0 & \label{eq:largest:45}\\
     & h_{oab} \in \{0,1\} & \forall o \in \mathcal{O}, a \in \mathcal{A}, b \in \mathcal{B}^0 & \label{eq:largest:46}\\
     %& \alpha_{oa} \in \{0,1\} & \forall o \in \mathcal{O}, a \in \mathcal{A} & \label{eq:largest:47}\\
     & s_o,\, u_{oa},\, v_{oa},\, w_{oa},\, z_{oa},\, \alpha_{oa},\, \beta_{oab} \in \{0,1\} & \forall o \in \mathcal{O}, a \in \mathcal{A}, b \in \mathcal{B} & \label{eq:largest:48}\\
     %& \gamma_{oa} \geq 0 & \forall o \in \mathcal{O}, a \in \mathcal{A} & \label{eq:largest:49}\\
     & u_o,\, G_{oa},\, \gamma_{oa},\, \delta_{oa} \geq 0 & \forall o \in \mathcal{O}, a \in \mathcal{A} & \label{eq:largest:50}
\end{align}

The first term of the objective function is the horizontal distance, up to the farthest aisle and back. The second term is the vertical distance to cross entirely the first and last aisles if at least two aisles are visited (represented by $s_o$). The last term is composed of two parts: the one related to variable $\gamma_{oa}$ is the vertical distance in the visited aisles that are neither the first, nor the last one visited. The term related to $\delta_{oa}$ is the vertical distance in case of a single aisle is visited (in this case a return policy is applied). Constraints~\eqref{eq:largest:01}-\eqref{eq:largest:17} are identical to the midpoint policy. Constraints~\eqref{eq:largest:18}-\eqref{eq:largest:19} ensure $g_{oa0}$ is computed correctly, i.e. they handle the border case of computing the gap between the front cross aisle and the closest pick. Constraints~\eqref{eq:largest:20}-\eqref{eq:largest:23} ensure $g_{oab}$ is computed correctly in the general case. Constraints~\eqref{eq:largest:24}-\eqref{eq:largest:26} ensure that the largest gap $G_{oa}$ is computed correctly. Constraints~\eqref{eq:largest:27}-\eqref{eq:largest:30} compute $\gamma_{oa} = s_ow_{oa}(\overline{b} + 1 - G_{oa})$. Constraints~\eqref{eq:largest:31}-\eqref{eq:largest:34} compute $\delta_{oa} = f_{oa}v_{oa}(1 - s_o)$. Constraints~\eqref{eq:largest:35}-\eqref{eq:largest:50} define the domains of the variables.

\end{document}